\documentclass[12pt,oneside,reqno]{amsart}
\usepackage{amssymb}
\usepackage{color}
\usepackage{tikz-cd}
\usetikzlibrary{decorations.markings}
\usepackage{xcolor}
\definecolor{deepgreen}{cmyk}{0.99998,0,1,0}

\usepackage{hyperref}
\hypersetup{hypertex=true,
	colorlinks=true,
	linkcolor=blue,
	anchorcolor=blue,
	citecolor=blue}
\usepackage{comment}
\usepackage{mathrsfs}

\newcommand{\cref}{\S\ \ref}
\newcommand{\Cref}{\S\ \ref}

\allowdisplaybreaks[4]

\usepackage{etoolbox}

  \usepackage{bm}
  \usepackage{bbm}
\theoremstyle{definition}
  \newtheorem{defi}{$\mathbf{Definition}$}[section]
  \newtheorem*{pro}{$\mathbf{Proof}$}
  
  \theoremstyle{plain}
  \newtheorem{theo}[defi]{$\mathbf{Theorem}$}
  \newtheorem{lemma}[defi]{$\mathbf{Lemma}$}
  \newtheorem{coro}[defi]{$\mathbf{Corollary}$}

  \newtheorem{prop}[defi]{$\mathbf{Proposition}$}
  
  \theoremstyle{remark}
  \newtheorem{remark}[defi]{$\mathbf{Remark}$}

\newcommand{\tro}{\mathrm{Tr}}

\newcommand{\pa}{\partial}

\newcommand{\lv}{\left\vert}
\newcommand{\rv}{\right\vert}
\newcommand{\lV}{\left\Vert}
\newcommand{\rV}{\right\Vert}

\newcommand{\bv}{\big\vert}

\newcommand{\bbv}{\bigg\vert}

\newcommand{\bV}{\big\Vert}

\newcommand{\bbV}{\bigg\Vert}

\newcommand{\h}{\mathcal{H}}
\newcommand{\op}{\mathrm{Op}}

\definecolor{pink}{RGB}{249,164,186}
\definecolor{grassgreen}{RGB}{128,255,0}

\numberwithin{equation}{section}

\title{Semiclassical analysis, Geometric representation and Quantum Ergodicity}
\author{Minghui Ma$^{1}$}
\address{$^1$Department of Mathematics and Statistics, University of New Hampshire}

\author{Qiaochu Ma$^{2}$}
\address{$^2$Department of Mathematics, The Pennsylvania State University}

\date{}

\makeatletter
\newcommand*{\rom}[1]{\expandafter\@slowromancap\romannumeral #1@}
\makeatother

\begin{document} 
\clearpage\maketitle
\begin{abstract}
	In this paper, we prove the equidistribution property of high-frequency eigensections of the Laplacian on a certain series of unitary flat bundles, using the mixture of semiclassical and geometric quantizations.
\end{abstract}

\section{Introduction}

\subsection{Backgrounds}\label{A1''}

The \emph{quantum ergodicity} (QE) was established by Shnirelman \cite{MR0402834}, Colin de Verdière \cite{MR818831} and Zelditch \cite{MR916129}, which states that on a compact Riemannian manifold whose geodesic flow is \emph{ergodic} with respect to the Liouville measure, the Laplacian has a \emph{density one} subsequence of eigenfunctions that tends to be \emph{equidistributed}.

Schrader-Taylor \cite{MR0995750} and Zelditch \cite{MR1183602} investigated QE on vector bundles. Jakobson-Strohmaier \cite{MR2276467} and Jakobson-Strohmaier-Zelditch \cite{MR2449142} considered the QE of some geometric differential operators, such as the Dirac operator and the Dolbeault Laplacian.

This paper aims to present the QE of \emph{unitary flat bundles}. Despite being locally trivial, their holonomy leads to intriguing global phenomena. In the following, we explain our main results in detail.

\subsection{Main results}

Let $(X,g^{TX})$ be an $m$-dimensional compact Riemannian manifold with the Levi-Civita connection $\nabla^{TX}$ and the volume form $dv_X$ induced by $g^{TX}$. Let $S^*X$ be the unit cotangent bundle of $X$, and let $(g_t)_{t\in\mathbb{R}}$ denote the geodesic flow on $S^*X$, obtained by identifying $T^*X$ with $TX$ through $g^{TX}$.

\subsubsection{The QE for one bundle}\label{A.2.1}

Let $U$ be a compact connected Lie group with Lie algebra $\mathfrak{u}$. Let $\rho$ be a $U$-representation of the fundamental group $\pi_1(X)$ of $X$,
\begin{equation}\label{1.1..}
\rho\colon \pi_1(X)\to U.
\end{equation}
Let $(V_\mu,h^{V_\mu})$ be an \emph{irreducible} unitary representation of $U$ with highest weight $\mu\in\mathfrak{u}^*$. Let $\widetilde{X}$ be the universal covering of $X$. We define
\begin{equation}\label{a1.}
	\begin{split}
F_\mu&=\widetilde{X}\times_{\pi_1(X)}V_\mu\\
&=\{(\widetilde{x},v)\in\widetilde{X}\times V_\mu\}/\big((\widetilde{x},v)\sim(\widetilde{x}\gamma,\rho(\gamma^{-1})v) \ \text{for any\ }\gamma\in\pi_1(X)\big).
	\end{split}
\end{equation}
Then $F_\mu$ is a flat vector bundle over $X$ with a natural flat connection $\nabla^{F_\mu}$, and $h^{V_\mu}$ induces a Hermitian metric $h^{F_\mu}$ on $F_\mu$, which is \emph{parallel} with respect to $\nabla^{F_\mu}$.

Let $\Delta^{F_\mu}$ be the nonnegative Laplacian acting on $\mathscr{C}^\infty(X,F_\mu)$. For a local orthonormal frame $\{e_i\}_{i=1}^m$ of $TX$, we have
\begin{equation}\label{a2.}
	\Delta^{F_\mu}=\sum_{i=1}^{m}-\big(\nabla^{F_\mu}_{e_i}\big)^2+\nabla^{F_\mu}_{\nabla^{TX}_{e_i}e_i}.
\end{equation}
We list all the eigenvalues $0\leqslant\lambda_{0}\leqslant\lambda_{1}\leqslant\cdots$ of $\Delta^{F_\mu}$ counted with multiplicity and the associated orthonormal eigensections:
\begin{equation}\label{a1}
	\Delta^{F_\mu}u_{j}=\lambda_{j}u_{j},\ \ \ \ \lV u_{j}\rV_{L^2(X,F_\mu)}=1.
\end{equation}

To simplify the notation, we shall write an additional overline on a trace operator or measure to signify the ``normalized" one in our subsequent discussion. For instance,
\begin{equation}
\overline{\tro}^{F_\mu}=\frac{1}{\dim F_\mu}\tro^{F_\mu},\ \ \ d\overline{v}_X=\frac{1}{\mathrm{Vol}(X)}dv_X.
\end{equation}
We now state the QE for $F_\mu$, see Theorem \ref{B2'} for the integrated version and Theorem \ref{B3'} for the full version with momentum variables.
\begin{theo}\label{A2}
Suppose that the geodesic flow $(g_t)_{t\in\mathbb{R}}$ on $S^*X$ is Anosov, and $\rho(\pi_1(X))\subset U$, given in \eqref{1.1..}, is dense, then there is a subset $\mathbb{B}\subseteq\mathbb{N}$ such that 
\begin{equation}\label{a5}
	\lim_{\lambda\to\infty}\frac{\lv\{j\in\mathbb{B}\mid \lambda_{j}\leqslant \lambda\}\rv}{\lv\{j\in\mathbb{N}\mid \lambda_{j}\leqslant \lambda\}\rv}=1,
\end{equation}
and for any $A\in\mathscr{C}^\infty(X,\mathrm{End}(F_\mu))$, we have
\begin{equation}\label{a6}
\lim_{\substack{j\in\mathbb{B},\\j\to\infty}}\langle Au_{j},u_{j}\rangle_{L^2(X,F_\mu)}={\int_X\overline{\tro}^{F_\mu}[A]d\overline{v}_X}.
\end{equation}
\end{theo}

\subsubsection{QE for a series of bundles}\label{1.2.2}
First, we introduce the general geometric setting.

Let $N$ be a compact complex manifold and $(L,h^L)$ a positive line bundle over $N$ with the Chern connection $\nabla^L$. Let $g^{T_\mathbb{R}N}$ be the Kähler metric on $T_\mathbb{R}N$ induced by the first Chern form $c_1(L,h^L)$ of $(L,h^L)$ and $dv_N$ the associated volume form. For $p\in\mathbb{N}^*$, let $H^{(0,0)}(N,L^p)$ be the space of holomorphic sections of $L^p=L^{\otimes p}$, the $p$-th tensor power of $L$, over $N$.  

Let $\mathcal{U}$ be the group of isomorphisms from $L$ to $L$ which preserve the Chern connection $\nabla^L$, each element of which restricts to a holomorphic isomorphism of $N$. Let 
\begin{equation}\label{1.8gg}
	\bm{\rho}\colon\pi_1(X)\to \mathcal{U}
\end{equation}
be a representation. We set
\begin{equation}\label{6}
	\mathscr{N}=\widetilde{X}\times_{\pi_1(X)}N,\ \ \ \mathscr{L}=\widetilde{X}\times_{\pi_1(X)}L,\ \ F_p=\widetilde{X}\times_{\pi_1(X)}H^{(0,0)}(N,L^p),
\end{equation}
then $\mathscr{N}$ is a flat $N$-bundle over $X$ with the volume form $dv_{\mathscr{N}}$ induced by $dv_X$ and $dv_N$, $\mathscr{L}$ is a line bundle on $\mathscr{N}$ with the metric $h^{\mathscr{L}}$ induced by $h^L$, and $F_p$ is a unitary flat bundle over $X$ with the natural flat connection $\nabla^{F_p}$ and the metric $h^{F_p}$ induced by $\langle\cdot,\cdot\rangle_{H^{(0,0)}(N,L^p)}$.

Let $\Delta^{F_p}$ the nonnegative Laplacian acting on $\mathscr{C}^\infty(X,F_p)$. We list eigenvalues $0\leqslant\lambda_{p,0}\leqslant\lambda_{p,1}\leqslant\cdots$ of $\Delta^{F_{p}}$ with multiplicity and associated orthonormal eigensections
\begin{equation}\label{1.8.'}
	\Delta^{F_{p}}u_{p,j}=\lambda_{p,j}u_{p,j},\ \ \ \ \lV u_{p,j}\rV_{L^2(X,F_{p})}^2=1.
\end{equation}
For
each $u_{p,j}\in \mathscr{C}^\infty(X,F_p)$ in \eqref{1.8.'}, by \eqref{6}, we can view $u_{p,j}\in\mathscr{C}^\infty(\mathscr{N},\mathscr{L}^p)$ and get the following set of probability measures on $\mathscr{N}$:
\begin{equation}\label{1.11gg}
\big\{\lV u_{p,j}(x,z)\rV_{\mathscr{L}^p}^2dv_{\mathscr{N}}(x,z)\big\}_{p,j\in\mathbb{N}}.
\end{equation}

We put natural projections $q\colon\mathscr{N}\to X$ and $\pi\colon T^*X\to X$ and let $q^*(T^*X)$ be the fibre product of $q$ and $\pi$. Recall that $S^*X$ is the unit cotangent bundle of $X$, then similarly we define $q^*(S^*X)$, and let us call it the \emph{augmented unit cotangent bundle}. Clearly
\begin{equation}\label{1.10..}
	q^*(T^*X)=T^*\widetilde{X}\times_{\pi_1(X)} N,\ \ \  q^*(S^*X)=S^*\widetilde{X}\times_{\pi_1(X)} N.
\end{equation}
Now we summarize all the geometric objects in the following diagram
\begin{equation}\label{fig1}
	\begin{tikzcd}[ampersand replacement=\&, column sep=normal,row sep=normal]
		\ \& \  \&		N\arrow[d,""swap]\arrow[dl,""swap]\arrow[dll,""swap] \& L^p\arrow[d,""swap]\arrow[l,""]  \\
		q^*(S^*X)\arrow[d,"q"swap]\arrow[r,""swap]	\& q^*(T^*X)\arrow[r,"\pi"swap]\arrow[d,"q"swap]  \&  \mathscr{N}\arrow[d,"q"swap] \& \mathscr{L}^p\arrow[l,""]\arrow[d,"Rq_*"]\\
		S^*X\arrow[r,""]	\&	T^*X \arrow[r,"\pi"] \& X \&F_p\arrow[l,""]
	\end{tikzcd}
\end{equation}
where $Rq_*$ represents the direct image.

\begin{defi}\label{A4}
	Let	the geodesic flow $(\widetilde{g_t})_{t\in\mathbb{R}}$ on $S^*\widetilde{X}$ act trivially on $N$. This action is $\pi_1(X)$-invariant, and it descents to a flow on $q^*(S^*X)$ by \eqref{1.10..}. We call it the \emph{horizontal geodesic flow} and denote it by $(\bm{g}_t)_{t\in\mathbb{R}}$.
\end{defi}
Note that globally $(\bm{g}_t)_{t\in\mathbb{R}}$ acts nontrivially along the fibre $N$ due to the holonomy. The Liouville measure on $S^*\widetilde{X}$ and $dv_N$ induce a measure $dv_{q^*(S^*X)}$ on $q^*(S^*X)$, and we call it the \emph{augmented Liouville measure}.

Now we can state our main result, uniform quantum ergodicity (UQE), on the equidis- tribution property of measures in \eqref{1.11gg}, see Theorem \ref{D9} for the integrated version and Theorem \ref{D10} for the full version involving momentum variables.
\begin{theo}\label{C9'}
	Suppose that the horizontal geodesic flow $(\bm{g}_t)_{t\in\mathbb{R}}$ on the augmented unit cotangent bundle $q^*(S^*X)$ is ergodic respect to the augmented Liouville measure $dv_{q^*(S^*X)}$, then there is a subset $\mathbb{B}\subseteq\mathbb{N}^2$ with the following uniform density one condition
	\begin{equation}\label{c22''}
		\lim_{\lambda\to\infty}\inf_{p\in\mathbb{N}}\frac{\lv\{(p,j)\in\mathbb{B}\mid\lambda_{p,j}\leqslant\lambda\}\rv}{\lv\{(p,j)\in\mathbb{N}^2\mid\lambda_{p,j}\leqslant\lambda\}\rv}=1,
	\end{equation}
	such that, for any $\mathscr{A}\in\mathscr{C}^\infty(\mathscr{N})$, we have the uniform limit
	\begin{equation}\label{9.}
		\lim_{\lambda\to\infty}\sup_{\substack{(p,j)\in\mathbb{B},\\
				\lambda_{p,j}\geqslant \lambda}}\lv\int_{\mathscr{N}}\mathscr{A}\lV u_{p,j}\rV_{\mathscr{L}^p}^2d{v}_{\mathscr{N}}-\int_{\mathscr{N}}\mathscr{A}d\overline{v}_{\mathscr{N}}\rv=0.
	\end{equation}
\end{theo}

Theorem \ref{C9'} shows that the quantum states tend to be equidistributed not only on the \emph{base manifold} $X$ but on the \emph{total space} $\mathscr{N}$ as well. In comparison with Theorem \ref{A2}, the two components in Theorem \ref{C9'}, namely the density one condition and the equidistribution result, are both \emph{uniform} with respect to $p\in\mathbb{N}$.

Now we discuss a special case of Theorem \ref{C9'}. We take $\bm{\rho}$ in \eqref{1.8gg} to be the representation $\rho$ given in \eqref{1.1..}, and  $(N,L)=(\mathcal{O}_\mu,L_\mu)$ in \eqref{6}, where $\mathcal{O}_\mu$ is the coadjoint orbit of a highest weight $\mu$ of $U$, and $L_\mu$ is the canonical line bundle over $\mathcal{O}_\mu$. Then we get
\begin{equation}\label{1.16gg}
	\begin{split}
&\big(\mathscr{N},q^*(T^*X),q^*(S^*X),\mathscr{L},F_p\big)\\
&=\big(\widetilde{X}\times_{\pi_1(X)}\mathcal{O}_\mu,T^*\widetilde{X}\times_{\pi_1(X)}\mathcal{O}_\mu,S^*\widetilde{X}\times_{\pi_1(X)}\mathcal{O}_\mu,\widetilde{X}\times_{\pi_1(X)}L_\mu,F_{p\mu}\big)
	\end{split}
\end{equation}
where $F_{p\mu}$ is defined in \eqref{a1.}, see \cref{Cd} for more details.

\begin{theo}\label{C9''}
	If the geodesic flow $(g_t)_{t\in\mathbb{R}}$ on $S^*X$ is Anosov, and $\rho(\pi_1(X))\subset U$, given in \eqref{1.1..}, is dense, then the dynamic assumption of Theorem \ref{C9'} is satisfied, making Theorem \ref{C9'} applicable for \eqref{1.16gg}.
\end{theo}

Note that if we apply Theorem \ref{A2} to each of $\{F_{p\mu}\}_{p\in\mathbb{N}}$, it would only imply a result weaker that Theorem \ref{C9'}. Specifically, we would replace $\lim_{\lambda\to\infty}\inf_{p\in\mathbb{N}}$ in \eqref{c22''} and $\lim_{\lambda\to\infty}\sup_{\substack{(p,j)\in\mathbb{B},\lambda_{p,j}\geqslant \lambda}}$ in \eqref{9.} with $\inf_{p\in\mathbb{N}}\lim_{\lambda\to\infty}$ and $\sup_{p\in\mathbb{N}}\lim_{(p,j)\in \mathbb{B},j\to\infty}$ respectively. This highlights the importance of \emph{uniformity} in Theorem \ref{C9'}.

\subsubsection{An example}\label{s1.2.3}

Let us explain Theorems \ref{A2}, \ref{C9'} and \ref{C9''} through an example. Now we take $U=\mathrm{SU}(2)$ in \eqref{1.1..} and $X=\Gamma_2\backslash \mathbb{H}^2$, a genus $2$ hyperbolic surface, where $\Gamma_2\subset\mathrm{PSL}(2,\mathbb{R})$ and $\Gamma_2\cong\{a_1,b_1,a_2,b_2\mid [a_1,b_1]\cdot[a_2,b_2]=1\}$. Note that in hyperbolic geometry, we usually use the left action, which is equivalent to the right one in \eqref{a1.} by setting $ x\gamma=\gamma^{-1}x$. We choose an \emph{irrational number} $\theta\in\mathbb{R}$ and set
\begin{equation}\label{0.17'}
	\rho(a_1)=\rho(b_1)=\begin{pmatrix}
		e^{-i\theta\pi/2}      & 0\\
		0    & e^{i\theta\pi/2}  
	\end{pmatrix},\ \ \rho(a_2)=\rho(b_2)=\begin{pmatrix}
		\cos \frac{\theta\pi}{2}&i\sin\frac{\theta\pi}{2} \\
		i\sin\frac{\theta\pi}{2}&\cos \frac{\theta\pi}{2}    
	\end{pmatrix},
\end{equation}
then $\rho$ extends to a representation $\rho\colon \Gamma_2\to\mathrm{SU}(2)$, and $\rho(\Gamma_2)\subset \mathrm{SU}(2)$ is dense.

The group $\mathrm{SU}(2)$ acts on $\mathbb{C}^2$ as well as its $p$-th symmetric power $\mathrm{Sym}^p(\mathbb{C}^2)$ for $p\in\mathbb{N}$, and these are all irreducible representations of $\mathrm{SU}(2)$. We take $\mu=1$ in \eqref{a1.}, then
\begin{equation}\label{1.7.}
	F_{p\mu}=\Gamma_2\backslash\big(\mathbb{H}^2\times\mathrm{Sym}^p(\mathbb{C}^2)\big)
\end{equation}
and the Laplacian $\Delta^{F_{p\mu}}$ is the \emph{Schrödinger-Pauli} spin $p/2$ operator, see Bolte-Glaser \cite{MR1794842} for a QE on Euclidean spaces with spin $1/2$. We list all eigenvalues $\{\lambda_{p,j}\}_{j\in\mathbb{N}}$ and their normalized eigensections $\{u_{p,j}\}_{j\in\mathbb{N}}$ as in \eqref{1.8.'}.

For each $p\in\mathbb{N}$ and $A\in\mathscr{C}^\infty\big(\Gamma_2\backslash \mathbb{H}^2,\Gamma_2\backslash\big(\mathbb{H}^2\times\mathrm{End}(\mathrm{Sym}^p\mathbb{C}^2)\big)\big)$, since $\rho(\Gamma_2)\subset \mathrm{SU}(2)$ in \eqref{0.17'} is dense, Theorem \ref{A2} reads as
\begin{equation}\label{a6}
	\lim_{\substack{j\in\mathbb{B},\\j\to\infty}}\langle Au_{p,j},u_{p,j}\rangle_{L^2(\Gamma_2\backslash \mathbb{H}^2,\Gamma_2\backslash(\mathbb{H}^2\times\mathrm{Sym}^p\mathbb{C}^2))}={\int_{\Gamma_2\backslash \mathbb{H}^2}\overline{\tro}^{\Gamma_2\backslash(\mathbb{H}^2\times\mathrm{Sym}^p(\mathbb{C}^2))}[A]d\overline{v}_{\Gamma_2\backslash \mathbb{H}^2}}.
\end{equation}

Now we turn to the the application of Theorem \ref{C9'}, we shall explain in an elementary way without giving $(L_\mu,\mathscr{L}_\mu)$. First, we reinterpret $\{u_{p,j}\}_{p,j\in\mathbb{N}}$ using geometric representation. Put $\mathbb{S}^3=\{(z_0,z_1)\in\mathbb{C}^2\mid\vert z_0\vert^2+\vert z_1\vert^2=1\}$ and $\mathbb{S}^1=\{\lambda\in\mathbb{C}\mid\vert \lambda\vert=1\}$, then the \emph{Hopf fibration} gives $\mathbb{S}^3/\mathbb{S}^1\cong \mathbb{CP}^1$. The Fubini-Study metric $\omega_{\mathrm{FS}}$ on $\mathbb{CP}^1$ is
\begin{equation}
	\omega_{\mathrm{FS}}=\frac{i}{2\pi}\frac{dz\wedge d\overline{z}}{(1+\vert z\vert^2)^2}
\end{equation}
when restricted to the open set $\{z_0\neq0\}$ with local coordinate $z=z_1/z_0$.

We can view $\mathrm{Sym}^p(\mathbb{C}^2)$ as the space of homogeneous polynomial functions of degree $p$ with two complex variables $(z_0,z_1)$:
\begin{equation}\label{1.12}
	\mathrm{Sym}^p(\mathbb{C}^2)=\{a_0z_0^p+a_1z_0^{p-1}z_1+\cdots+a_p{z_1}^p\mid a_0,\cdots,a_p\in\mathbb{C}\},
\end{equation}
and for $g\in\mathrm{SU}(2), s\in \mathrm{Sym}^p(\mathbb{C}^2)$, the action $g\cdot s$ is given by
\begin{equation}\label{1.19g}
	(g\cdot s)(z_0,z_1)=s(g^{-1}\cdot(z_0,z_1)).
\end{equation}
For $s,s'\in \mathrm{Sym}^p(\mathbb{C}^2)$, the function 
\begin{equation}\label{1.20g}
s\cdot\overline{s'}\colon (z_0,z_1)\in\mathbb{S}^3\mapsto s(z_0,z_1)\cdot\overline{s'(z_0,z_1)}
\end{equation}
is $\mathbb{S}^1$-invariant, which descents to a smooth function on $\mathbb{CP}^1$. Then we can define the $L^2$-product on $\mathrm{Sym}^p(\mathbb{C}^2)$ by
\begin{equation}\label{1.13}
	\langle s,s'\rangle_{\mathrm{Sym}^p(\mathbb{C}^2)}=\int_{\mathbb{CP}^1}s(z)\cdot\overline{s'(z)}\omega_{\mathrm{FS}}(z),
\end{equation}
and give the following orthonormal basis 
\begin{equation}\label{1.13'}
	\big\{\big((p+1)\tbinom{p}{j}\big)^{1/2}z_0^{p-j}z_1^{j}\big\}_{0\leqslant j\leqslant p}.
\end{equation}

Let $\mathscr{C}^\infty_{\Gamma_2}(\mathbb{H}^2,\mathrm{Sym}^p(\mathbb{C}^2))$ denote the space of smooth $\Gamma_2$-invariant $\mathrm{Sym}^p(\mathbb{C}^2)$-valued functions on $\mathbb{H}^2$, in other words, functions $u\in\mathscr{C}^\infty(\mathbb{H}^2,\mathrm{Sym}^p(\mathbb{C}^2))$ such that 
\begin{equation}\label{1.23g}
	u(\gamma \widetilde{x})=\rho(\gamma)u(\widetilde{x})
\end{equation}
for any $\gamma\in\Gamma_2$ and $\widetilde{x}\in \mathbb{H}^2$. Clearly we have the following isomorphism
\begin{equation}\label{1.24g}
\mathscr{C}^\infty\big(\Gamma_2\backslash \mathbb{H}^2,\Gamma_2\backslash\big(\mathbb{H}^2\times\mathrm{Sym}^p(\mathbb{C}^2)\big)\big)\cong \mathscr{C}^\infty_{\Gamma_2}\big(\mathbb{H}^2,\mathrm{Sym}^p(\mathbb{C}^2)\big).
\end{equation}
For any eigensection $u_{p,j}\in\mathscr{C}^\infty(\Gamma_2\backslash \mathbb{H}^2,\Gamma_2\backslash(\mathbb{H}^2\times\mathrm{Sym}^p(\mathbb{C}^2)))$, by \eqref{1.24g} we view $u_{p,j}\in\mathscr{C}^\infty_{\Gamma_2}(\mathbb{H}^2,\mathrm{Sym}^p(\mathbb{C}^2))$, then by \eqref{1.20g} we can define the following function on $\mathbb{H}^2\times\mathbb{CP}^1$:
\begin{equation}\label{1.25g}
(\widetilde{x},z)\in \mathbb{H}^2\times\mathbb{CP}^1\mapsto \lv u_{p,j}(\widetilde{x})(z)\rv_\mathbb{C}^2.
\end{equation}
From \eqref{1.19g} and \eqref{1.23g}, the function given in \eqref{1.25g} is also $\Gamma_2$-invariant, that is,
\begin{equation}
\lv u_{p,j}(\gamma\widetilde{x})(\rho(\gamma)z)\rv_\mathbb{C}^2=\lv\big(\rho(\gamma)u_{p,j}(\widetilde{x})\big)(\rho(\gamma)z)\rv_\mathbb{C}^2=\lv u_{p,j}(\widetilde{x})(z)\rv_\mathbb{C}^2.
\end{equation}

In \eqref{1.16gg}, we take
\begin{equation}\label{1.15..}
(X,\mathcal{O}_\mu,\mathscr{N})=\big(\Gamma_2\backslash\mathbb{H}^2,\mathbb{CP}^1,\Gamma_2\backslash\big(\mathbb{H}^2\times \mathbb{CP}^1\big)\big),
\end{equation} 
where $\Gamma_2$ acts on $\mathbb{CP}^1$ through \eqref{0.17'}. Then the function in \eqref{1.25g} passes to a function in $\mathscr{C}^\infty(\Gamma_2\backslash(\mathbb{H}^2\times \mathbb{CP}^1))$, and we denote it by $\lv u_{p,j}\rv_{\mathbb{C}}^2$. Let $dv_{\Gamma_2\backslash(\mathbb{H}^2\times \mathbb{CP}^1)}$ denote the volume form on $\Gamma_2\backslash(\mathbb{H}^2\times \mathbb{CP}^1)$ locally given by $dv_{\Gamma_2\backslash\mathbb{H}^2}(x)\omega_{\mathrm{FS}}(z)$.

Finally, we get probability measures $\{\lv u_{p,j}\rv_\mathbb{C}^2d{v}_{\Gamma_2\backslash(\mathbb{H}^2\times \mathbb{CP}^1)}\}_{p,j\in\mathbb{N}}$ on $\Gamma_2\backslash(\mathbb{H}^2\times \mathbb{CP}^1)$ corresponding to \eqref{1.11gg}, locally written as $\big\{\lv u_{p,j}(x)(z)\rv_\mathbb{C}^2dv_{\Gamma_2\backslash\mathbb{H}^2}(x)\omega_{\mathrm{FS}}(z)\big\}_{p,j\in\mathbb{N}}$. Recall that $\rho(\Gamma_2)\subset \mathrm{SU}(2)$ in \eqref{0.17'} is dense, hence Theorems \ref{C9'} and \ref{C9''} read as
	\begin{equation}\label{1.31g}
		\begin{split}
	\lim_{\lambda\to\infty}\sup_{\substack{(p,j)\in\mathbb{B},\\
		\lambda_{p,j}\geqslant \lambda}}\bbv\int_{\Gamma_2\backslash(\mathbb{H}^2\times \mathbb{CP}^1)}\mathscr{A}\lv u_{p,j}\rv_\mathbb{C}^2&d{v}_{\Gamma_2\backslash(\mathbb{H}^2\times \mathbb{CP}^1)}\\
&-\int_{\Gamma_2\backslash(\mathbb{H}^2\times \mathbb{CP}^1)}\mathscr{A}d\overline{v}_{\Gamma_2\backslash(\mathbb{H}^2\times \mathbb{CP}^1)}\bbv=0.
		\end{split}
\end{equation}

We note that \eqref{1.31g} is nontrivial, since functions given in \eqref{1.13'} are \emph{not} automatically equidistributed on $\mathbb{CP}^1$. This is an easy consequence of the Stirling formula applied to the projection of $\{(z_0,z_1)\in\mathbb{S}^3\mid\lv z_0\rv\leqslant 1/2\}$ to $\mathbb{CP}^1$, which gives
\begin{equation}
	\lim_{p\to\infty}\sup_{\substack{p/3\leqslant j\leqslant 2p/3,\\
			\{(z_0,z_1)\in\mathbb{S}^3\mid\lv z_0\rv\leqslant 1/2\}}}\big((p+1)\tbinom{p}{j}\big)^{1/2}\lv z_0^{p-j}z_1^{j}\rv=0.
\end{equation}
Therefore, if \eqref{0.17'} is replaced with the \emph{trivial representation}, \eqref{1.31g} fails.

\subsection{Main technique}

The proof of Theorem \ref{A2} is rather classical, while the key technique leading to Theorem \ref{C9'} is the \emph{``mixed quantization"}, see \cref{Cc}, which brings together semiclassical and geometric quantizations.

Let us explain through the example discussed in \cref{s1.2.3}. Consider the tautological line bundle $\mathcal{O}(-1)$ over $\mathbb{CP}^1$ given by
\begin{equation}
	\mathcal{O}(-1)=\{(z,v)\in\mathbb{CP}^1\times \mathbb{C}^2\mid v\in z\},
\end{equation}
which has a metric $h^{\mathcal{O}(-1)}$ induced by $\langle\cdot,\cdot\rangle_{\mathbb{C}^2}$. We set $\mathcal{O}(1)=\mathcal{O}(-1)^*$ and $\mathcal{O}(p)=\mathcal{O}(1)^{\otimes p}$, the $p$-th tensor power of $\mathcal{O}(1)$ for $p\in\mathbb{N}$, with a metric $h^{\mathcal{O}(p)}$ given by $h^{\mathcal{O}(-1)}$, then $c_1(\mathcal{O}(1),h^{\mathcal{O}(1)})=\omega_{\mathrm{FS}}$. Let $H^{(0,0)}(\mathbb{CP}^1,\mathcal{O}(p))$ denote the set of holomorphic sections of $\mathcal{O}(p)$. Each $s\in\mathrm{Sym}^p(\mathbb{C}^2)$ corresponds to a function on $\mathcal{O}(-1)$ by $(z,v)\mapsto s(v)$, and this gives an isomorphism \begin{equation}\label{1.23.}
\mathrm{Sym}^p(\mathbb{C}^2)\cong H^{(0,0)}(\mathbb{CP}^1,\mathcal{O}(p)).
\end{equation}
Then the $L^2$-metric $\langle\cdot,\cdot\rangle_{L^2}$ on $H^{(0,0)}(\mathbb{CP}^1,\mathcal{O}(p))$ induced by $(h^{\mathcal{O}(p)},\omega_{\mathrm{FS}})$ is just the metric $\langle\cdot,\cdot\rangle_{\mathrm{Sym}^p(\mathbb{C}^2)}$ in \eqref{1.13} through \eqref{1.23.}.

For $\mathscr{A}\in \mathscr{C}^\infty(\Gamma_2\backslash(\mathbb{H}^2\times \mathbb{CP}^1))$, we set $T_{\mathscr{A},p}\in\mathscr{C}^\infty(\Gamma_2\backslash\mathbb{H}^2,\mathrm{End}(\Gamma_2\backslash(\mathbb{H}^2\times\mathrm{Sym}^p\mathbb{C}^2)))$ such that for each $x\in \Gamma_2\backslash\mathbb{H}^2$ and $v,w\in \Gamma_2\backslash(\mathbb{H}^2\times\mathrm{Sym}^p\mathbb{C}^2)|_{x}\cong H^{(0,0)}(\mathbb{CP}^1,\mathcal{O}(p))$,
\begin{equation}\label{0.17}
	\langle T_{\mathscr{A},p}(x)v,w\rangle_{L^2(\mathbb{CP}^1,\mathcal{O}(p))}=\int_{\mathbb{CP}^1}\mathscr{A}(x,z)\langle v(z),w(z)\rangle_{h^{\mathcal{O}(p)}}\omega_{\mathrm{FS}}(z).
\end{equation}
Locally we compute the following term in \eqref{1.31g} that
\begin{equation}\label{0.25}
	\begin{split}
		\int_{\Gamma_2\backslash(\mathbb{H}^2\times \mathbb{CP}^1)}&\mathscr{A}\lv u_{p,j}\rv^2_{\mathbb{C}}dv_{\Gamma_2\backslash(\mathbb{H}^2\times \mathbb{CP}^1)}\\
		&=\int_{\Gamma_2\backslash\mathbb{H}^2}\Big(\int_{\mathbb{CP}^1}\mathscr{A}(x,z)\bv u_{p,j}(x)(z)\bv^2_{h^{\mathcal{O}(p)}}\omega_{\mathrm{FS}}(z)\Big)dv_{\Gamma_2\backslash\mathbb{H}^2}(x)\\
		&=\int_{\Gamma_2\backslash\mathbb{H}^2}\langle T_{\mathscr{A},p}(x)u_{p,j}(x),u_{p,j}(x)\rangle_{\Gamma_2\backslash(\mathbb{H}^2\times\mathrm{Sym}^p\mathbb{C}^2)} dv_{\Gamma_2\backslash\mathbb{H}^2}(x)\\
		&=\langle\op_h(T_{\mathscr{A},p})u_{p,j},u_{p,j}\rangle_{L^2(\Gamma_2\backslash\mathbb{H}^2,\Gamma_2\backslash(\mathbb{H}^2\times\mathrm{Sym}^p\mathbb{C}^2))}.
	\end{split}
\end{equation}
In \eqref{0.17}, $T_{\mathscr{A},p}$ is indeed the \emph{Berezin-Toeplitz quantization} of $\mathscr{A}$ along the fibre $\mathbb{CP}^1$, and in \eqref{0.25}, $\op_h(\cdot)$ is the \emph{Weyl quantization}, even though $T_{\mathscr{A},p}(x)u_{p,j}(x)$ here is the direct product without momentum variables.

This explains the relevance of our results to two quantizations: \emph{high-frequency eigensections are governed by the Weyl quantization and the behavior of an infinite number of linear spaces are regulated by the Berezin-Toeplitz quantization. Combining them enables simultaneous control of the high-frequency eigensections of an infinite number of bundles}.

\subsection{Asymptotic torsion, QE and UQE}\label{S1.4.}
Our motivation stems from the study of the torsion invariant of flat bundles. Let $(F,\nabla^F)$ be a flat bundle over $X$ with metric $h^F$, Ray-Singer \cite{MR295381} defined their \emph{analytic torsion} $T(X,F)$ as an analogue of the \emph{Reidemeister-Franz torsion} \cite{MR3069647,MR1581473}. These two torsions coincide for unitary flat bundles by Cheeger \cite{MR528965} and Müller \cite{MR498252}. Bismut-Zhang \cite{MR1185803} generalized this to arbitrary flat vector bundles.

As a global spectral invariant, the analytic torsion is difficult to calculate explicitly in general. Bergeron-Venkatesh \cite{MR3028790} discussed the asymptotics of analytic torsions of quotients of symmetric spaces by a decreasing sequence of lattices in an underlying Lie group, and studied the growth of torsion elements in the homology of an arithmetic group. For a compact $3$-dimensional hyperbolic manifold $X=\Gamma\backslash \mathbb{H}^3$ where $\Gamma\subset \mathrm{SL}(2,\mathbb{C})$ and $\bm{F}_p=\Gamma\backslash(\mathbb{H}^3\times\mathrm{Sym}^p(\mathbb{C}^2))$, Müller \cite{MR3220447} obtained $\lim_{p\to\infty}p^{-2}T(X,\bm{F}_p)=(4\pi)^{-1}\mathrm{Vol}(X)$ using the Selberg trace formula. In \cite{MR2838248,MR3615411}, Bismut-Ma-Zhang gave a general construction of a family of nonunitarily flat vector bundles $\{\bm{F}_p\}_{p\in\mathbb{N}}$ on any compact manifold and they computed the leading term in the asymptotics of analytic torsions.

Then it seems natural to ask, in the cases where the above asymptotic torsion results do not cover, like unitary flat bundles over a Riemannian surface discussed in \cref{s1.2.3}, can we find other asymptotic spectral information? This motivates us to consider the equidistribution property of eigensections.

We are inspired by the framework of \cite{MR3615411} in various aspects. For instance, $\{F_{p}\}_{p\in\mathbb{N}}$ in \eqref{6} is an analogue of their construction of nonunitary $\{\bm{F}_p\}_{p\in\mathbb{N}}$ in \cite[\S\,4.1]{MR3615411}, and they use differential operators with coefficients in Berezin-Toeplitz quantization in \cite[\S\,9.8]{MR3615411}. This was further developed by Ma \cite{MR4665497} for full asymptotic torsions. We should emphasize that the setup of \cite{MR3615411} is very general, the vector bundles can be induced by a \emph{general fibration} which is \emph{not} necessarily a principal bundle, nor flat, see the asymptotic holomorphic torsions of Puchol \cite{MR4611826} for an example.

After Shnirelman's original contribution, there have been many QE-type results in different settings, for example, on manifolds with boundary by Zelditch-Zworski \cite{MR1372814}, on noncompact manifolds by Zelditch \cite{MR1105653}, on moduli spaces by
Baskin-Gell-Redman-Han \cite{MR4584856}. The large-scale QE was considered by Le Masson-Sahlsten \cite{MR3732880} on manifolds and by Anantharaman-Le Masson \cite{MR3322309} on regular graphs, which might be compared with Bergeron-Venkatesh \cite{MR3028790}. For a detailed overview of results on QE and related subjects, we refer to Anantharaman \cite{MR4477342}, Dyatlov \cite{MR4396066} and Sarnak \cite{MR2774090}. As we discussed, there are already some results in analytic torsion and the equidistribution property that are analogous to each other, so we can also expect to study the QE and UQE of vector bundles in more diverse settings.

\subsection{Comparison with previous work}

For a principal $U$ bundle $P_U\to X$, which is not necessarily flat, using the Fourier integral operator technique, Schrader-Taylor \cite{MR0995750} and Zelditch \cite{MR1183602} investigated QE on the ladder subspace of $L^2(P_U)$, which is essentially equivalent to the QE of vector bundles $F_{p\mu}=P_U \times_U V_{p\mu}$ for a highest weight $\mu$ of $U$.

However, there are trade-offs to this nonflat consideration. Firstly, if we list eigenvalues and eigensections as in \eqref{1.8.'}, their results are limited to the case where $\lambda_{p,j}$ and $p^2$ are approximately the same size, in other word, $0<c\leqslant\lambda_{p,j}/p^2\leqslant C<\infty$ for some $C,c>0$, in contrast to Theorem \ref{C9'} where UQE is independent of $p$. The key point is that we need to analyze the Berezin-Toeplitz quantization using symmetry and then obtain an estimate of the quantum variance without the $O(p^{-1})$ term in the remainder, see \cref{Sdf} for more details. Secondly, their dynamic assumptions in \cite[Theorem 9.1]{MR0995750} and \cite[(0.1)]{MR1183602} are difficult to describe geometrically, and practical examples to which their assumptions hold are hard to come by, see discussions in \cite[\S\,8]{MR0995750} and \cite[(3.20)]{MR1183602}.

\subsection{Organization of the paper}

This paper is organized as follows. In \Cref{B}, we give some fundamental tools for the semiclassical analysis of unitary flat bundles. In \cref{B'}, we prove the general version of Theorem \ref{A2}. In \cref{sM}, we introduce the mixed quantization and prove the full version of Theorem \ref{C9'}. In \cref{D}, we detail more examples to which the main results are applicable and discuss the set of semiclassical measures.

\subsection{Notations}
In the whole paper, for $\alpha=(\alpha_1,\cdots,\alpha_k)\in\mathbb{N}^k,x=(x_1,\cdots,x_k)\in\mathbb{R}^k$, we denote $\vert\alpha\vert=\sum_{i=1}^k\alpha_i$, $x^\alpha=x_1^{\alpha_1}\cdots x_k^{\alpha_k},\pa^\alpha_x=\frac{\pa^{\alpha_1}}{\pa x_1^{\alpha_1}}\cdots\frac{\pa^{\alpha_k}}{\pa x_k^{\alpha_k}}$.

\subsection*{Acknowledgment} 

We would like to express our gratitude to Xiaonan Ma and Stephane Nonnenmacher for careful reading and many corrections for earlier versions of this manuscript. We thank Dennis Sullivan for suggesting the dense image condition in our main results. We are grateful to Yves Coudène, Yulin Gong, Antonin Guilloux, Long Jin, Dave Witte Morris, Shu Shen and Akshay Venkatesh for helpful discussions and suggestions. We express gratitude to the anonymous referees for valuable comments, especially for bringing \cite{MR0995750} and \cite{MR1183602} to our attention. Q. M. would like to thank Nigel Higson for hospitality and was supported by the NSF grant DMS-1952669.

\subsection*{Declarations}
The authors have no relevant financial or financial interests to disclose. Data sharing does not apply to this article as no datasets were generated or analyzed during the current study.

\section{Semiclassical Analysis on Unitary Flat Bundles}\label{B}

In this section, we provide some results on the semiclassical analysis of unitary flat bundles, for more details we refer to Dyatlov-Zworski \cite{MR3969938} and Zworski \cite{MR2952218}.

This section is organized as follows. In \cref{Ba}, we recall some basic facts on the Weyl quantization. In \cref{Bb}, we review the local Weyl law of the Laplacian. In \cref{Bc}, we give the Egorov theorem, which links the classical and quantum evolutions.

\subsection{Weyl quantization}\label{Ba}

Let us \emph{fix} a \emph{finite good cover} $\{U_\alpha\}$ of $X$ by coordinate charts as in \cite[Theorem 5.1]{MR658304}, demanding any finite intersection $U_{\alpha_{1}}\cap\cdots\cap U_{\alpha _{k}}$ to be \emph{differentiably contractible}. Then for any unitary flat vector bundle $(F,\nabla^F)$ on $X$  with a (flat) Hermitian metric $h^F$, it is induced by a representation $\rho\colon\pi_1(X)\to\mathrm{U}_{\dim F}$,
\begin{equation}\label{2.1}
F=\widetilde{X}\times_{\pi_1(X)}\mathbb{C}^{\dim F}.
\end{equation}
We can choose local sections $\{v_{\alpha,i}\in\mathscr{C}^\infty(U_\alpha,F)\}_{i=1}^{\dim F}$ on each $U_\alpha$ such that $\nabla^Fv_{\alpha,i}=0$ and $\langle v_{\alpha,i},v_{\alpha,j}\rangle_{h^F}=\delta_{i,j}$. This gives a local trivialization $F|_{U_\alpha}\cong \mathbb{C}^{\dim F}$ that the transition functions $\phi_{\alpha,\beta}\in\mathscr{C}^\infty(U_\alpha\cap U_\beta,\mathrm{U}_{\dim F})$ are indeed \emph{constant}. Let $\lV\cdot\rV_F$ be the norm on $F$ induced by $h^F$ and $\lV\cdot\rV_{\mathrm{End}(F)}$ the associated \emph{operator norm} on $\mathrm{End}(F)$.

Let $\{\phi_\alpha\}$ be a partition of unity with respect to $\{U_\alpha\}$. Then for any $k\in\mathbb{Z}$, we set the semiclassical Sobolev space $H_h^k(X,F)$ with norm $\lV\cdot\rV_{H_h^k(X,F)}$: for any $0<h\leqslant1$ and $s\in\mathscr{C}^\infty(X,F)$,
\begin{equation}
	\lV s\rV_{H_h^k(X,F)}^2=\begin{cases}
		\sum_{\alpha}\sum_{\lv\beta\rv\leqslant k}h^{2\lv\beta\rv}\lV\pa_x^\beta(\phi_\alpha s)\rV_{L^2(U_\alpha,F)}^2,\ \ \text{if\ }k\in\mathbb{N},\\
		\sup_{\lV s'\rV_{H^{-k}_h(X,F)}=1}\lv\langle s,s'\rangle_{L^2(X,F)}\rv^2,\ \ \text{if\ }k\in\mathbb{Z}\backslash\mathbb{N}.
	\end{cases}
\end{equation}
Note that these are all standard Sobolev spaces, but the norms depend on $h$.

Note that we have a natural projection $\pi\colon T^*X\to X$. For $k\in\mathbb{Z}$, we say that a smooth section $A(x,\xi)\in\mathscr{C}^\infty(T^*X,\pi^*\mathrm{End}(F))$ is in the symbol class $S_F^k$ if and only if for each $j\in\mathbb{N}$, the following Kohn-Nirenberg norm is finite:
\begin{equation}\label{b2}
	\lv A\rv^{(k)}_j=\max_{\begin{subarray}{c}
			\lv\beta\rv,\lv\gamma\rv\leqslant j,\\
			U_\alpha\in\{U_\alpha\}
	\end{subarray}}\sup_{x\in U_\alpha,\xi\in T^*_xX}\langle\xi\rangle^{-k+\lv\gamma\rv}\bV\partial_x^{\beta}\partial_{\xi}^{\gamma}A(x,\xi)\bV_{\mathrm{End}(F_x)}.
\end{equation}
We set $S_F^{-\infty}=\bigcap\limits_{k\in\mathbb{Z}}S_F^{k}$. 

We choose a set of nonnegative smooth functions $\{\phi'_\alpha\}$ such that $\mathrm{supp}(\phi'_\alpha)\subset U_\alpha$ and $\phi_\alpha'\equiv1$ on an open set containing $\mathrm{supp}(\phi_\alpha)$.
For $A\in S_F^k$, its quantization, is defined as \begin{equation}\label{a.2}
	\op_h(A)=\sum_\alpha \phi_\alpha' \op_h(\phi_\alpha A) \phi_\alpha'\ \ \text{for\ }0<h\leqslant1,
\end{equation}
where $\op_h(\phi_\alpha A)$ is the \emph{Weyl quantization} on $\mathbb{R}^m$ \cite[(4.1.1)]{MR2952218}:
\begin{equation}\label{a.5}
\big(\op_h(\phi_\alpha A)s\big)(x)=\frac{1}{(2\pi h)^m}\int_{\mathbb{R}^{m}}\int_{\mathbb{R}^{m}}e^{\frac{i}{h}\langle x-y,\xi\rangle}\big(\phi_\alpha A\big)(\textstyle{\frac{x+y}{2}},\xi)s(y)dyd\xi.
\end{equation}
\begin{remark}\label{B.1}
	Before getting to the main text, we note that the proof of the estimates in this section proceeds just as the case $(F,\nabla^F,h^F)=(\mathbb{C},d,\langle\cdot,\cdot\rangle_{\mathbb{C}})$, the \emph{trivial line bundle}, the only point that calls for special mention is that we can choose \emph{uniform constants} in all inequalities, namely that they are \emph{independent} of the unitary flat bundle $(F,\nabla^F,h^F)$ and only depend on $\{U_\alpha\}$ and $\{\phi_\alpha,\phi_\alpha'\}$. This can be seen from the trivialization argument at the beginning of \cref{Ba}. On each $U_\alpha$, we can always view $(F,\nabla^F,h^F)$ as the \emph{standard trivial unitary bundle} $(\mathbb{C}^{\dim F},d,\langle\cdot,\cdot\rangle_{\mathbb{C}^{\dim F}})$, so we only need to prove the uniformity for trivial bundles, which is usually easy to verify by replacing $\lv s(y)\rv_{\mathbb{C}}$ and $\lv A(x,\xi)\rv_{\mathbb{C}}$ with $\lV s(y)\rV_{F}$ and $\lV A(x,\xi)\rV_{\text{End}(F)}$ respectively. This observation is vitally essential, for we shall deal with an infinite number of bundles simultaneously in \cref{sM}.
	
	Moreover, due to the necessity for uniformity, we shall write the quantitative remainder terms for the estimates in this section. Although these terms are not explicitly stated in \cite{MR2952218}, they are indeed included in the proofs therein, see more discussions in \cref{S2.4.}.
\end{remark}

Now we list some standard facts on the Weyl quantization. The integral in \eqref{a.5} should be viewed as a map between Sobolev spaces, indeed, for $k,\ell\in\mathbb{Z}$, there are $C>0,j\in\mathbb{N}$ such that for any $A\in S^k_F$ and $0<h\leqslant1$, we have
\begin{equation}\label{b11}
	\lV\mathrm{Op}_{h}(A)s\rV_{H_h^{\ell}(X,F)}\leqslant C\lv A\rv_{j}^{(k)}\lV s\rV_{H^{\ell+k}_h(X,F)},
\end{equation}
see \cite[Proposition E.19]{MR3969938} and \cite[Theorem 4.23]{MR2952218}. If $A\in S_F^0$, the adjoint of its quantization with respect to the $L^2$-metric is given by
\begin{equation}\label{b12}
	\mathrm{Op}_{h}(A)^*=\mathrm{Op}_{h}(A^*).
\end{equation}
For $A\in S_F^{k_1}$ and $B\in S_F^{k_2}$, set 
\begin{equation}\label{b7..}
A*_0B=AB,\ \ A*_1B=\frac{1}{2i}\big({\pa_{\xi_j}}A\cdot{\pa_{x_j}}B-{\pa_{x_j}}A\cdot{\pa_{\xi_j}}B\big).
\end{equation}
Then we have the product formula \cite[Theorem 9.5]{MR2952218} with a remainder term
\begin{equation}\label{b7}
	\mathrm{Op}_{h}(A)\mathrm{Op}_{h}(B)=\mathrm{Op}_{h}(A*_0B)+h\mathrm{Op}_h(A*_1B)+\frac{h^2}{2}R_{h,2}(A,B),
\end{equation}
and for any $\ell\in\mathbb{Z}$, there are $C>0,k\in\mathbb{N}$ such that for any $0<h\leqslant1$,
\begin{equation}
\lV R_{h,2}(A,B)s\rV_{H^{\ell}_h(X,F)}\leqslant C\lv A\rv_{k}^{(k_1)}\lv B\rv_{k}^{(k_2)}\lV s\rV_{H^{\ell+k_1+k_2-2}_h(X,F)}.
\end{equation}

Next, we have some useful estimates from the local structure of the Laplacian, see \cite[Theorems 14.3, 14.6]{MR2952218}.
\begin{prop}\label{B2}
For any $\ell\in\mathbb{Z}$, there is $C>0$ such that for any $0<h\leqslant1$,
	\begin{equation}\label{b14}
	 \lV\big(\op_h(\lv\xi\rv_{T^*X}^2)-h^2\Delta^F\big)s\rV_{H^{\ell}_h(X,F)}\leqslant Ch\lV s\rV_{H_h^{\ell+1}(X,F)}.
	\end{equation}
For any $\ell\in\mathbb{N}$, there is $C>0$ such that for any $z\in\mathbb{C}\backslash\mathbb{R}$, 
\begin{equation}\label{2.12g}
\lV(h^2\Delta^F-z)^{-1}s\rV_{H^{\ell+1}_h(X,F)}\leqslant C\big(1+\lv\mathrm{Im}(z)\rv^{-\ell-1}\big)\lV s\rV_{H_h^\ell(X,F)}.
\end{equation}
\end{prop}

Let $\mathscr{S}(\mathbb{R})$ be the Schwartz space with a series of norms $\{\lv\cdot\rv_\ell\}_{\ell\in\mathbb{N}}$ given by
\begin{equation}
	\lv\phi\rv_{\ell}=\max_{0\leqslant i,j\leqslant \ell}\sup_{t\in\mathbb{R}}\lv t^i\pa_t^j\phi(t)\rv_\mathbb{C}.
\end{equation}

By Proposition \ref{B2}, Helffer-Sjöstrand formula \cite[Theorem 11.8]{MR2952218} and Beals theorem \cite[Theorem 9.12]{MR2952218}, we get the following result on the functional calculus of the Laplacian, see \cite[Theorem 14.9]{MR2952218}.
\begin{theo}\label{B3.}
For any $i\in\mathbb{N}$, there are $C>0,\ell\in\mathbb{N}$ such that for any $0<h\leqslant 1$ and $\phi\in\mathscr{S}(\mathbb{R})$, we have
\begin{equation}\label{b15}
	\lV\big(\op_h\big(\phi(\lv\xi\rv_{T^*X}^2)\big)-\phi(h^2\Delta^F)\big)s\rV_{H^{i}_h(X,F)}\leqslant Ch\lv\phi\rv_{\ell}\lV s\rV_{L^2(X,F)}.
\end{equation}
\end{theo}
Note that in the proof of Beals theorem, replacing $b(x,\xi)\in\mathscr{S}'(\mathbb{R}^m)$ in \cite[(8.1.1)]{MR2952218} with $\langle A(x,\xi)v,w\rangle_{\mathbb{C}^{\dim F}}$ for some $A\in \mathscr{S}'(\mathbb{R}^{2m},\mathbb{C}^{\dim F})$ and $v,w\in \mathbb{C}^{\dim F}$, one gets the corresponding inequality for all flat bundles with the same constants as the trivial line bundle, which also confirms Remark \ref{B.1}.

\subsection{Local Weyl law}\label{Bb}

Let $\omega_{T^*X}$ be the canonical symplectic form on $T^*X$, and we denote by $dv_{T^*X}$ the associated volume form on $T^*X$, that is,
\begin{equation}
\omega_{T^*X}=\sum_{i=1}^{m}d\xi_i\wedge dx_i,\ \ dv_{T^*X}=\frac{\omega_{T^*X}^m}{m!}.
\end{equation}

We first state a classical trace formula.
\begin{lemma}\label{B3}
	For any $0<h\leqslant1$ and $A\in S_F^{-\infty}$, the operator $\op_h(A)$ is of trace class on $L^2(X,F)$, and its trace is given by
	\begin{equation}\label{b.14}
		\frac{(2\pi h)^m}{\dim F}\tro\big[\op_h(A)\big]=\int_{T^*X}\overline{\tro}^{\pi^*F}[A]dv_{T^*X}.
	\end{equation}
In general, there is $C>0$ such that for any linear operator $T_h\colon L^2(X,F)\to H^{{m}+1}_h(X,F)$ bounded uniformly for $0<h\leqslant1$, it is of trace class on $L^2(X,F)$ and
\begin{equation}\label{b.16}
	\frac{(2\pi h)^m}{\dim F}\lv\mathrm{Tr}[T_h]\rv\leqslant C\lV T_h\rV_{L^2(X,F)\to H^{{m}+1}_h(X,F)}.
\end{equation}
\end{lemma}

As \cite[Theorem 15.3]{MR2952218}, by \eqref{b11}, \eqref{b7}, \eqref{b15}, \eqref{b.14}, \eqref{b.16} and an obvious inequality 
\begin{equation}\label{b18}
\lv\overline{\tro}^F[T]\rv\leqslant \lV T\rV_{\mathrm{End}(F)}
\end{equation}
for any $T\in\mathrm{End}(F)$, we have the following local Weyl law.
\begin{theo}\label{B4}
	There are $C>0,i,\ell\in\mathbb{N}$ such that for any $0<h\leqslant1,\phi\in\mathscr{S}(\mathbb{R})$ and $A\in S_F^0$, we have
	\begin{equation}\label{b.18}
			\bbv\frac{(2\pi h)^m}{\dim F}\tro\big[\phi(h^2\Delta^F)\op_h(A)\big]-\int_{T^*X}\phi(\lv\xi\rv_{T^*X}^2)\overline{\tro}^F[A]dv_{T^*X}\bbv\leqslant Ch\lv\phi\rv_{\ell}\lv A\rv^{(0)}_i.
	\end{equation}
\end{theo}

Let $\{\lambda_j\}_{j\in\mathbb{N}}$ be the eigenvalues of $\Delta^F$ with associated orthonormal eigensections $\{u_j\}_{j\in\mathbb{N}}$ as in \eqref{a1}. Taking $A\equiv1$ in Theorem \ref{B4} and approximating the function $\mathbbm{1}_{[0,1]}$ from above and below by functions in $\mathscr{S}(\mathbb{R})$, we get the following Weyl law.
\begin{coro}\label{B5}
	For any $\varepsilon>0$, there is $h_\varepsilon>0$ such that for any $0<h\leqslant h_\varepsilon$,
	\begin{equation}\label{b.19.}
		\bbv\frac{(2\pi h)^m}{\dim F}\lv\{j\mid\lambda_j\leqslant h^{-2}\}\rv-\frac{1}{m}\cdot\mathrm{Vol}(S^*X)\bbv\leqslant\varepsilon.
	\end{equation}
\end{coro}
For an eigenvalue $\lambda_j$ of $\Delta^F$, we set $h_j=\lambda_j^{-\frac{1}{2}}$. By \eqref{b11}, \eqref{b7} and \eqref{b14}, we obtain the following lemma concerning the concentration of eigensections on $S^*X$.
\begin{lemma}\label{B6}
There are $C>0,i\in\mathbb{N},$ such that for any $A\in S_F^0$ with $A|_{S^*X}=0$ and $0<h\leqslant1$, we have
	\begin{equation}\label{b.20}
		\sup_{\lambda_j\geqslant h^{-2}}\bV\op_{h_j}(A)u_j\bV_{L^2(X,F)}\leqslant Ch\lv A\rv^{(0)}_i.
	\end{equation}
\end{lemma}

Following the proof of \cite[Lemma 2.1]{MR4396066}, we replace $A(x,\xi)$ in \eqref{b.18} with $A'(x,\xi)$ such that $A'\big(x,a\xi\big)=A(x,\xi)$ for  $(x,\xi)\in S^*X$ and $a^2\in\mathrm{supp}(\phi)$. Then by \eqref{b.19.} and \eqref{b.20}, we obtain the following result.
\begin{coro}\label{B7}
	There is $k\in\mathbb{N}$ such that for any $\phi\in\mathscr{S}(\mathbb{R})$ with $\mathrm{supp}(\phi)\subset (0,\infty)$, there is $C>0$ such that for any $A\in S_F^0$ and $0<h\leqslant 1$, we have
	\begin{equation}\label{b.21}
		\begin{split}
			\bbv\frac{(2\pi h)^m}{\dim F}\sum_{j}\phi(h^2\lambda_j)\langle \op_{h_j}(A)u_j,u_j\rangle-\int_{T^*X}\phi(\lv\xi\rv_{T^*X}^2)\overline{\tro}^F\big[A\big(x,\lv\xi\rv_{T^*X}^{-1}\xi\big)\big]dv_{T^*X}\bbv\\
			\leqslant C h\lv A\rv^{(0)}_k.
		\end{split}
	\end{equation}
\end{coro}

\subsection{Egorov theorem}\label{Bc}

Let $(\psi_t)_{t\in\mathbb{R}}$ be the \emph{Hamiltonian flow} of $\lv\xi\rv^2_{T^*X}/2$ with respect to $\omega_{T^*X}$, which restricts to the \emph{geodesic flow} $(g_t)_{t\in\mathbb{R}}$ on $S^*X$. For $t\in\mathbb{R}$ and $A\in \mathscr{C}^\infty(T^*X,\pi^*\mathrm{End}(F))$, we can define the action of the morphism $\psi_t$ on $A$: for $(x,\xi)\in T^*X$, let $\gamma\colon[0,\lv t\rv]\to T^*X$ be a smooth curve given by $\gamma(s)=\psi_{t-\mathrm{sign}(t)s}(x,\xi)$ for $0\leqslant s\leqslant \lv t\rv$, which connects $\psi_t(x,\xi)$ and $(x,\xi)$, then we let $(\psi_t\cdot A)(x,\xi)$ be the parallel transport of $A(\psi_t(x,\xi))$ along $\gamma$ with respect to the connection $\nabla^{\pi^*\mathrm{End}(F)}$.

By \eqref{2.1}, each $A\in \mathscr{C}^\infty(T^*X,\pi^*\mathrm{End}(F))$ corresponds to a $\pi_1(X)$-invariant $\widetilde{A}\in\mathscr{C}^\infty(T^*\widetilde{X},\mathrm{End}(\mathbb{C}^{\dim F}))$:
\begin{equation}\label{2.22}
\widetilde{A}\big(\gamma\cdot(\widetilde{x},\widetilde{\xi})\big)=\rho(\gamma)\widetilde{A}(\widetilde{x},\widetilde{\xi})\rho(\gamma)^{-1},
\end{equation}
where $(\widetilde{x},\widetilde{\xi})\in T^*\widetilde{X}$ and $\gamma\cdot(\widetilde{x},\widetilde{\xi})=(\gamma\cdot\widetilde{x},(\gamma^{-1})^*\widetilde{\xi})$ for $\gamma\in\pi_1(X)$. We call $\widetilde{A}$ the \emph{lifting} of $A$. The flow ${\psi}_t$ also lifts to the \emph{Hamiltonian flow} of $\bv\widetilde{\xi}\bv^2_{T^*\widetilde{X}}/2$ with respect to $\omega_{T^*\widetilde{X}}$. Then $(\psi_t\cdot A)$ can be equivalently described at the level of lifting by
\begin{equation}\label{2.23}
\widetilde{\psi}_t\cdot \widetilde{A}=\widetilde{\psi_t\cdot A}\ \ \ \text{for}\ (\widetilde{\psi}_t\cdot \widetilde{A})(\widetilde{x},\widetilde{\xi})=\widetilde{A}(\widetilde{\psi}_t(\widetilde{x},\widetilde{\xi})).
\end{equation}

We define the \emph{Schrödinger propagator} $U^{F}_{t,h}$ of $\Delta^{F}$ by
\begin{equation}\label{b23}
	U^{F}_{t,h}=\exp({-ith\Delta^{F}}/2)\ \ \text{for\ } t\in\mathbb{R},0<h\leqslant1. 
\end{equation}
By \eqref{b7..}, \eqref{b7} and \eqref{b14}, the operator $U^{F}_{t,h}$ quantizes the flow $(\psi_t)_{t\in\mathbb{R}}$ as made precisely by the following Egorov theorem \cite[Theorem 15.2]{MR2952218}.
\begin{theo}\label{B9}
There are $j,k\in\mathbb{N}$ such that for any $T>0$, there is $C>0$ such that for any $0<h\leqslant 1,0\leqslant t\leqslant T$ and $A\in S_F^{-\infty}$, we have
	\begin{equation}\label{b23.}
		\lV\big(U^{F}_{-t,h}\mathrm{Op}_{h}(A)U^{F}_{t,h}-\op_h(\psi_t\cdot A)\big)s\rV_{L^2(X,F)}\leqslant Ch\lv A\rv^{(j)}_{k}\lV s\rV_{L^2(X,F)}.
	\end{equation}
\end{theo}

For $A\in \mathscr{C}^\infty(T^*X,\pi^*\mathrm{End}(F))$, define its $L^2$-norm restricted to $S^*X$ by
\begin{equation}\label{b25}
\lV A\rV_{L^2(S^*X,\pi^*\mathrm{End}(F))}^2=\int_{S^*X}\overline{\tro}^F[A^*A]d\overline{v}_{S^*X},
\end{equation}
and for any $T\in\mathbb{R}$, its time average $\langle A\rangle_T\in\mathscr{C}^\infty(T^*X,\pi^*\mathrm{End}(F))$ is defined by
\begin{equation}\label{b25'}
	\langle A\rangle_T=\frac{1}{T}\int_{0}^{T}(\psi_t\cdot A)dt.
\end{equation}

Now we state the main result of this section, an estimate of quantum variance.
\begin{theo}\label{B10}
There is $k\in\mathbb{N}$ such that for any $0<a<b<\infty$ and $0<T<\infty$, there exist $C_{a,b}, C_{a,b,T}>0$ such that for any $A\in S_F^{0}$ and $0<h\leqslant 1$, we have
	\begin{equation}\label{b26}
		\begin{split}
			&\frac{(2\pi h)^m}{\dim F}\sum_{ah^{-2}\leqslant\lambda_j\leqslant bh^{-2}}\bbv\langle \op_{h_j}(A)u_j,u_j\rangle_{L^2(X,F)}-\int_{S^*X}\overline{\tro}^{\pi^*F}[A]d\overline{v}_{S^*X}\bbv^2\\
			&\leqslant C_{a,b}\bbV\langle A\rangle_T-\int_{S^*X}\overline{\tro}^{\pi^*F}[A]d\overline{v}_{S^*X}\cdot\mathrm{Id}_{\pi^*F}\bbV^2_{L^2(S^*X,\pi^*\mathrm{End}(F))}+C_{a,b,T}h\big(\lv A\rv^{(0)}_k\big)^2.
		\end{split}
	\end{equation}
\end{theo}

\begin{pro}
From \eqref{b.20}, we can assume without loss of generality that $A\in S^{-\infty}_F$. By \eqref{b23}, we have $U^F_{t,h_j}u_j=e^{-ith_j\lambda_j/2}u_j$, therefore,
\begin{equation}\label{b27}
	\big\langle U^{F}_{-t,h_j}\op_{h_j}(A)U^{F}_{t,h_j}u_j,u_j\big\rangle=\big\langle \op_{h_j}(A)u_j,u_j\big\rangle.
\end{equation}
In Corollary \ref{B7}, we take a nonnegative $\phi\in\mathscr{S}(\mathbb{R})$ with $\mathrm{supp}(\phi)\subset(0,\infty)$ and  $\phi\equiv1$ on $[a,b]$, then \eqref{b26} follows immediately from \eqref{b11}, \eqref{b7}, \eqref{b.16}, \eqref{b23.} and \eqref{b27}. \qed
\end{pro}

\subsection{Integer constants}\label{S2.4.}

To determine the integer constants in the estimates of this section, we have two starting points. Without loss of generality we work on $X=\mathbb{R}^m$ and $F=\mathbb{C}$ in this subsection. 

The first starting point is that, there is $C>0$ such that for any $A\in S^0$ and $0<h\leqslant 1$,
\begin{equation}\label{2.31.}
\lV\op_h(A)s\rV_{L^2}\leqslant C\lv A\rv^{(0)}_{14m+4}\lV s\rV_{L^2},
\end{equation}
see the proof of \cite[Theorem 4.23]{MR2952218}. The second starting point is the product formula \cite[Theorems 4.11, 9.5]{MR2952218}, that is, for any $A\in S^{k_1}$ and $B\in S^{k_1}$, we have a $A\#_hB\in S^{k_1+k_2}$ such that
\begin{equation}
\op_h(A)\op_h(B)=\op_h(A\#_hB).
\end{equation} 
Moreover, there exist $A*_kB\in S^{k_1+k_2-k}$ and $R_{k,h}(A,B)\in S^{k_1+k_2-k}$ such that for any $k\in\mathbb{N}$, we have an asymptotic expansion
\begin{equation}
A\#_hB=\sum_{i=0}^{k-1}h^iA*_iB+h^{k}R_{k,h}(A,B),\ \ \ A\#_hB=R_{0,h}(A,B)
\end{equation}
where $A*_0B$ and $A*_1B$ are given in \eqref{b7..}. Also, for any $j\in\mathbb{N}$, there is $C>0$ such that
\begin{equation}\label{2.33}
\lv R_{k,h}(A,B)\rv^{(k_1+k_2-k)}_j\leqslant C\lv A\rv^{(k_1)}_{4m+1+2k+\lv k_1\rv+\lv k_2\rv+2j}\lv B\rv^{(k_1)}_{4m+1+2k+\lv k_1\rv+\lv k_2\rv+2j}.
\end{equation}
We note that the subscripts on the right side of \eqref{2.33} include $m$. This arises because $R_{k,h}(A,B)$ is defined by an \emph{oscillatory integral}. To ensure convergence, higher-order derivatives are required to induce decay in the integrand, necessitating the involvement of $m$, see the proof of \cite[Theorem 4.17]{MR2952218}.

In principle, all other integer constants in this section can be derived from Proposition \ref{B2}, \eqref{2.31.} and \eqref{2.33}. However, determining these constants is complex, particularly due to the subscripts in \eqref{2.33}.  For our purpose, acknowledging their existence is sufficient. 

For example, \eqref{b11} can be reduced to \eqref{2.31.} using the product formula \eqref{2.33}, see \cite[Theorem E.19]{MR3969938}. In \eqref{b23.}, we need to control $\lv\psi_t\cdot A\rv^{(0)}_{k}$ for a $k\in\mathbb{N}$ by \eqref{2.31.} and \eqref{2.33}. For $T>0$, there is $C>0$, such that for any $0\leqslant t\leqslant T$,
\begin{equation}
\lv\psi_t\cdot A\rv^{(0)}_{k}\leqslant C\lv A\rv^{(-k)}_{k},
\end{equation}
where the superscript on the right is $-k$ because the pushforward of momentum derivatives $(\psi_t)_*\pa_{\xi_j}$ have position derivatives component $\pa_{x_i}$, see \cite[Remark 15.2]{MR2952218}.

\section{The QE for $F_\mu$}\label{B'}

In this section, we present the full version of Theorem \ref{A2}, the quantum ergodicity (QE) for $F_\mu$.

This section is organized as follows. In \cref{Ba'}, we recall some basic properties of the Anosov flow following \cite{MR1377265}. In \cref{Ba''}, we prove an $L^2$-ergodic theorem for the geodesic flow action on $L^2(S^*X, \pi^*\mathrm{End}(F_\mu))$. In \cref{Bb'}, we prove the QE for $F_\mu$.

\subsection{The Anosov flow}\label{Ba'}

For $(x,\xi)\in S^*X$, we define the \emph{stable} $W_{S^*X}^s(x,\xi)$ and \emph{unstable} manifolds $W_{S^*X}^u(x,\xi)$ by
\begin{equation}\label{3.1}
	\begin{split}
		W_{S^*X}^s(x,\xi)&=\{(x',\xi')\in S^*X\mid d(g_t(x,\xi),g_t(x',\xi'))\to0\ \ \mathrm{as\ }t\to\infty\},\\
		W_{S^*X}^u(x,\xi)&=\{(x',\xi')\in S^*X\mid d(g_t(x,\xi),g_t(x',\xi'))\to0\ \ \mathrm{as\ }t\to-\infty\}.
	\end{split}
\end{equation}
Let $\mathcal{X}$ be the infinitesimal generator of the geodesic flow on $S^*X$, for $(x,\xi)\in S^*X$,
\begin{equation}\label{2.2.}
\mathcal{X}_{(x,\xi)}=\frac{d}{dt}\Big|_{t=0}g_t\cdot(x,\xi).
\end{equation}

\begin{defi}
 We say that the geodesic flow on $S^*X$ is \emph{Anosov} if there exist constants $C,c>0$ and a $(g_t)_*$-invariant continuous splitting
\begin{equation}
T(S^*X)=\mathbb{R}\mathcal{X}\oplus E_{S^*X}^s\oplus E_{S^*X}^u
\end{equation}
such that for  $t\geqslant0,(x,\xi)\in S^*X,v_s\in E_{S^*X}^s(x,\xi)$ and $v_u\in E_{S^*X}^u(x,\xi)$, we have
\begin{equation}
	\begin{split}
\lv(g_t)_*v_s\rv\leqslant Ce^{-ct}\lv v_s\rv,\ \ \ 
\lv(g_{-t})_*v_u\rv\leqslant Ce^{-ct}\lv v_u\rv.
	\end{split}
\end{equation}
\end{defi}
If the geodesic flow on $S^*X$ is Anosov, the tangent distributions of foliations $W^s_{S^*X}$ and $W^u_{S^*X}$ are precisely $E^s_{S^*X}$ and $E^u_{S^*X}$. Moreover, the Anosov property is well preserved when lifting to the universal covering $\widetilde{X}$.

\subsection{The Hopf argument}\label{Ba''}

Recall that $\widetilde{X}$ is the universal covering of $X$, then we have a Riemannian metric $g^{T\widetilde{X}}$ on $\widetilde{X}$ induced by $g^{TX}$. Let $S^*\widetilde{X}$ be the unit cotangent bundle of $\widetilde{X}$, then the geodesic flow $(g_t)_{t\in\mathbb{R}}$ on $S^*X$ lifts to the geodesic flow $(\widetilde{g}_t)_{t\in\mathbb{R}}$ on $S^*\widetilde{X}$.

For a measurable section $A$ of $\pi^*\mathrm{End}(F_\mu)$ on $S^*X$, as in \eqref{2.22} and \eqref{2.23}, its lifting $\widetilde{A}$ is a measurable $\pi_1(X)$-invariant $\mathrm{End}(V_\mu)$-valued function on $S^*\widetilde{X}$, and the $(g_t)_{t\in\mathbb{R}}$-action on $A$ is equivalent to the $(\widetilde{g}_t)_{t\in\mathbb{R}}$-action on its lifting
\begin{equation}\label{b1'}
	\big(\widetilde{g}_t\cdot \widetilde{A}\big)(\widetilde{x},\widetilde{\xi})=\widetilde{A}\big(\widetilde{g}_t\cdot(\widetilde{x},\widetilde{\xi})\big).
\end{equation}
Similar to \eqref{b25'}, the time average $\langle \widetilde{A}\rangle_T$ of $\widetilde{A}$ is defined by
\begin{equation}
\langle \widetilde{A}\rangle_T=\frac{1}{T}\int_{0}^{T}\widetilde{g}_t\cdot \widetilde{A}dt.
\end{equation} 

We say that $A$ is $(g_t)_{t\in\mathbb{R}}$-invariant if $\{(x,\xi)\in S^*X\mid(g_t\cdot A)(x,\xi)\neq A(x,\xi)\}$ is of $0$-measure for every $t\in\mathbb{R}$, since $\pi_1(X)$ is \emph{countable}, this equivalently means that the lifting $\widetilde{A}$ is $(\widetilde{g}_t)_{t\in\mathbb{R}}$-invariant.

\begin{prop}\label{B1'}
Under the assumption as in Theorem \ref{A2}. If $A\in L^2(S^*X,\pi^*\mathrm{End}(F_\mu))$ is $(g_t)_{t\in\mathbb{R}}$-invariant, then there is $c\in\mathbb{C}$ that $A=c\cdot\mathrm{Id}_{\pi^*F_\mu}$. Moreover, for any $A\in L^2(S^*X,\pi^*\mathrm{End}(F_\mu))$, we have
\begin{equation}\label{b3'}
\lim_{T\to\pm\infty}\bbV\langle A\rangle_T-\Big(\int_{S^*X}\overline{\tro}^{\pi^*F_\mu}[A]d\overline{v}_{S^*X}\Big)\mathrm{Id}_{\pi^*F_\mu}\bbV_{L^2(S^*X,\pi^*\mathrm{End}(F_\mu))}^2=0.
\end{equation}
\end{prop}

\begin{pro}
Proceeding as in \cite[Proposition 2.6]{MR1377265}, we only need to prove \eqref{b3'} when $A\in\mathscr{C}(S^*X,\pi^*\mathrm{End}(F_\mu))$. Since $dv_{S^*X}$ is $(g_t)_{t\in\mathbb{R}}$-invariant and $h^{F_\mu}$ is parallel with respect to $\nabla^{F_\mu}$, by the von Neumann theorem for Hilbert spaces \cite[Theorem 3.1.6]{MR3558990}, there are $A^\pm\in L^2(S^*X,\pi^*\mathrm{End}(F_\mu))$ such that $\lim_{T\to\pm\infty}\lV\langle A\rangle_T-A^\pm\rV_{L^2(S^*X,\pi^*\mathrm{End}(F_\mu))}^2=0$ and $A^+=A^-$ almost everywhere. Then we can find a sequence $\{T_\ell\}_{\ell\in\mathbb{N}}$ such that
\begin{equation}\label{3.8'}
\lim_{\ell\to\infty}T_\ell=\infty,\ \ \lim_{\ell\to\infty}\langle A\rangle_{\pm T_\ell}=A^\pm
\end{equation}
almost everywhere. Since $A$ is continuous on the compact space $S^*X$, it is indeed \emph{uniformly continuous}. Given that $U$ acts isomorphically on $(V_\mu,h^{V_\mu})$, the lifting $\widetilde{A}$ of $A$ is also \emph{uniformly continuous} on $S^*\widetilde{X}$. Therefore, by \eqref{3.1} and \eqref{3.8'}, we see that $\widetilde{A^+}(\widetilde{x},\widetilde{\xi})=\widetilde{A^+}(\widetilde{x}',\widetilde{\xi}')$ whenever $(\widetilde{x}',\widetilde{\xi}')\in W^s_{S^*\widetilde{X}}(\widetilde{x},\widetilde{\xi})$ and $\widetilde{A^+}(\widetilde{x},\widetilde{\xi})$ is defined. Similarly, $\widetilde{A^-}(\widetilde{x},\widetilde{\xi})=\widetilde{A^-}(\widetilde{x}',\widetilde{\xi}')$ whenever $(\widetilde{x}',\widetilde{\xi}')\in W^u_{S^*\widetilde{X}}(\widetilde{x},\widetilde{\xi})$ and $\widetilde{A^-}(\widetilde{x},\widetilde{\xi})$ is defined. As the geodesic flow $(\widetilde{g}_t)_{t\in\mathbb{R}}$ is Anosov, $W^s_{S^*\widetilde{X}},W^u_{S^*\widetilde{X}}$ and $(\widetilde{g}_t)_{t\in\mathbb{R}}$ are transversally absolutely continuous foliations, along which $\widetilde{A^+}=\widetilde{A^-}$ is constant, then by \cite[Proposition 3.12]{MR1377265}, $\widetilde{A^+}$ is a constant $\mathrm{End}(V_\mu)$-valued function on $S^*\widetilde{X}$ almost everywhere. Since $\widetilde{A^+}$ is $\pi_1(X)$-invariant and the image of $\rho\colon\pi_1(X)\to U$ is dense, $\widetilde{A^+}$ commutes with the $U$-action on $V_\mu$. By the Schur lemma, $\widetilde{A^+}=c\mathrm{Id}_{V_\mu}$ for a $c\in\mathbb{C}$, then $A^+=c\mathrm{Id}_{\pi^*F_\mu}$ and it remains to determine $c$. By the $(g_t)_{t\in\mathbb{R}}$-invariance of $dv_{S^*X}$ and the definition of parallel transport, for any $T\in\mathbb{R}$,
\begin{equation}
	\int_{S^*X}\overline{\tro}^{\pi^*F_\mu}[\langle A\rangle_T]d\overline{v}_{S^*X}=\int_{S^*X}\overline{\tro}^{\pi^*F_\mu}[A]d\overline{v}_{S^*X}.
\end{equation}
Then by taking $T=T_\ell$ and let $\ell\to\infty$, we get
\begin{equation}
\int_{S^*X}\overline{\tro}^{\pi^*F_\mu}[A^+]d\overline{v}_{S^*X}=\int_{S^*X}\overline{\tro}^{\pi^*F_\mu}[c\mathrm{Id}_{\pi^*F_\mu}]d\overline{v}_{S^*X}=c,
\end{equation}
which gives \eqref{b3'}.\qed
\end{pro}

\subsection{Full versions of Theorem \ref{A2}}\label{Bb'}

Now we state the main result of this section.
\begin{theo}\label{B2'}
Under the assumption as in Theorem \ref{A2}, for any $0<a<b<\infty$ and $A\in S^{0}_{F_\mu}$, we have
	\begin{equation}\label{b5'}
		\begin{split}
		\lim_{h\to 0}\frac{(2\pi h)^m}{\dim F_\mu}\sum_{ah^{-2}\leqslant\lambda_j\leqslant bh^{-2}}\bbv\langle \op_{h_j}(A)u_j,u_j\rangle-\int_{S^*X}\overline{\tro}^{\pi^*F_\mu}[A]d\overline{v}_{S^*X}\bbv^2=0.
		\end{split}
	\end{equation}
\end{theo}

\begin{pro}
	Taking $\lim_{T\to\infty}\lim_{h\to0}$ in \eqref{b26}, this together with \eqref{b3'} implies \eqref{b5'}.\qed
\end{pro}

By taking $a=1,b=4$ in Theorem \ref{B2'}, we get the following result based on a well-known diagonal extraction argument.
\begin{theo}\label{B3'}
	There is a subset $\mathbb{B}\subseteq\mathbb{N}$ satisfying \eqref{a5} such that for any $A\in S^0_{F_\mu}$,
	\begin{equation}\label{3.10}
		\lim_{\substack{j\in\mathbb{B},\\ j\to\infty}}\langle\op_{h_{j}}(A)u_{j},u_{j}\rangle_{L^2(X,F_\mu)}=\int_{S^*X}\overline{\tro}^{\pi^*F_\mu}[A]d\overline{v}_{S^*X}.
	\end{equation}
\end{theo}

\section{Mixed quantization}\label{sM}

In this section, we introduce the mixed quantization and present the uniform quantum ergodicity (UQE) associated with the unitary flat vector bundles $\{F_{p}\}_{p\in\mathbb{N}}$.

This section is organized as follows. In \cref{Ca}, we review the Berezin-Toeplitz quantization formalism. In \cref{Cc}, we introduce the mixed quantization procedure, which involves the Weyl quantization along the base manifold and the fibrewise Berezin-Toeplitz quantization. In \cref{Ce}, we provide a criterion for the ergodicity of horizontal geodesic flow. In \cref{Sdf}, we prove the UQE. In \cref{Cd}, we prove Theorem \ref{C9''} using the Borel-Weil-Bott theorem.

\subsection{The Berezin-Toeplitz quantization}\label{Ca}

In this subsection, we describe the Berezin-Toeplitz quantization introduced by Berezin \cite{MR0395610} and Boutet de Monvel-Guillemin \cite{MR620794}, and further developed by Bordemann-Meinrenken-Schlichenmaier \cite{MR1301849}, Schlichenmaier \cite{MR1805922} and Ma-Marinescu \cite{MR2339952,MR2393271}.

Let $(N,J)$ be a compact complex manifold with $\dim_{\mathbb{C}}N=n$ (complex dimension) and $(L,h^L)$ a \emph{positive} holomorphic line bundle on $N$. The first Chern form $c_1(L,h^L)$ gives a Kähler metric $g^{T_\mathbb{R}N}$ on $T_\mathbb{R}N$ and a volume form $dv_N$ on $N$. 

For ${p\in{\mathbb{N}}}$, put $L^p=L^{\otimes p}$, the $p$-th tensor power of $L$ and $H^{(0,0)}(N,L^p)$ the space of holomorphic sections of $L^p$ on $N$. Let $\langle\cdot,\cdot\rangle_{H^{(0,0)}(N,L^p)}$ be the $L^2$-product on $H^{(0,0)}(N,L^p)$ induced by $(dv_{N},h^{L})$ and $P_p\colon L^2(N,L^p)\rightarrow H^{(0,0)}(N,L^p)$
the associated orthogonal projection. Then $P_p$ has a smooth kernel $P_p(z,z')$ such that
\begin{equation}\label{4.1..}
(P_ps)(z)=\int_NP_p(z,z')s(z')dv_N(z'),
\end{equation}
and it is called the \emph{Bergman kernel}.

\begin{defi}\label{Cb2}
	The Berezin-Toeplitz quantization of $\mathcal{H}\in \mathscr{C}^\infty(N)$ is a sequence of linear operators $\{T_{\mathcal{H},p}\in\mathrm{End}(L^2(N,L^p))\}_{p\in{\mathbb{N}}}$ given by $T_{\mathcal{H},p}=P_p\mathcal{H}P_p$, in other words, for any $s,s'\in H^{(0,0)}(N,L^p)$, we have
	\begin{equation}\label{c3}
\langle T_{\h,p}s,s'\rangle_{L^2(N,L^p)}=\int_{N}\h(z)\langle s(z),s'(z)\rangle_{h^{L^p}}dv_{N}(z).
	\end{equation}
\end{defi}

By \eqref{c3}, clearly we have
\begin{equation}\label{c4}
	\lV T_{\h,p}\rV_{\mathrm{End}(H^{(0,0)}(N,L^p))}\leqslant \lv \h\rv_{\mathscr{C}^0(N)}.
\end{equation}

The operator $T_{\mathcal{H},p}$ has a smooth kernel $T_{\mathcal{H},p}(z,z')$. On the diagonal, $T_{\mathcal{H},p}(z,z)\in \mathrm{End}(L^p)=\mathbb{C}$. We have the following expansion with a uniform estimate of the remainder from the proof of \cite[Lemma 7.2.4]{MR2339952}.
\begin{lemma}\label{Cb3}
There exists $C>0$ such that for any ${p\in{\mathbb{N}}}$ and $\h\in\mathscr{C}^\infty(N)$, 
	\begin{equation}\label{cb4}
		\bv p^{-n}T_{\mathcal{H},p}(z,z)-\mathcal{H}(z)\bv_{\mathscr{C}^0(N)}\leqslant C\lv\mathcal{H}\rv_{\mathscr{C}^{2}(N)}p^{-1}.
	\end{equation}
In particular, taking the integral on both sides of \eqref{cb4}, we get an asymptotic Hirzebruch-Riemann-Roch theorem \cite[Theorem 1.4.6]{MR2339952}
	\begin{equation}\label{cb5}
		\begin{split}
\dim H^{(0,0)}(N,L^p)=\mathrm{Vol}(N)p^n+\mathcal{O}(p^{n-1}),
		\end{split}
	\end{equation}
	 and a trace formula
	 	\begin{equation}\label{cb5'}
	 	\begin{split}
	 	\lv\overline{\tro}^{H^{(0,0)}(N,L^p)}[T_{\mathcal{H},p}]-\int_{N}\mathcal{H}d\overline{v}_{N}\rv\leqslant C\lv\mathcal{H}\rv_{\mathscr{C}^{2}(N)}p^{-1}.
	 	\end{split}
	 \end{equation}
\end{lemma}

\begin{lemma}
	For any ${p\in{\mathbb{N}}}$, $\h\in\mathscr{C}^\infty(N)$ and $s\in L^2(N,L^p)$, we have
\begin{equation}\label{4.6''}
	\langle T_{\mathcal{H},p}^*T_{\mathcal{H},p}s,s\rangle_{L^2(N,L^p)}\leqslant \langle T_{\lv\mathcal{H}\rv_{\mathbb{C}}^2,p}s,s\rangle_{L^2(N,L^p)},
\end{equation}
where $\lv\mathcal{H}\rv_{\mathbb{C}}^2\in\mathscr{C}^\infty(N)$ is defined by $\lv\mathcal{H}\rv_{\mathbb{C}}^2\colon z\in N\mapsto \lv\mathcal{H}(z)\rv_{\mathbb{C}}^2$. In particular,
\begin{equation}\label{4.7''}
	\overline{\tro}^{H^{(0,0)}(N,L^p)}\big[T_{\mathcal{H},p}^*T_{\mathcal{H},p}\big]\leqslant \overline{\tro}^{H^{(0,0)}(N,L^p)}\big[T_{\lv\mathcal{H}\rv_{\mathbb{C}}^2,p}\big]
\end{equation}
\end{lemma}

\begin{proof}
Writing the obvious inequality 
\begin{equation}
\lV P_p\mathcal{H}P_ps\rV_{L^2(N,L^p)}^2\leqslant\lV\mathcal{H}P_ps\rV_{L^2(N,L^p)}^2
\end{equation}into the language of \eqref{c3}, we obtain \eqref{4.6''}.
\end{proof}

\begin{remark}
Indeed, we have the following more precise product formula
	\begin{equation}\label{cb7}
		\bV T_{\mathcal{H},p}T_{\mathcal{H}',p}-T_{\mathcal{H}\mathcal{H}',p}\bV_{\mathrm{End}(H^{(0,0)}(N,L^p))}\leqslant C\lv\mathcal{H}\rv_{\mathscr{C}^{2}(N)}\cdot\lv\mathcal{H}'\rv_{\mathscr{C}^{2}(N)}p^{-1},
	\end{equation}
see \cite[Theorem 7.4.1]{MR2339952}, while we only need the much weaker result \eqref{4.6''}.
\end{remark}

\subsection{The mixed quantization on $S^k(q^*(T^*X))$}\label{Cc}

For $k\in\mathbb{Z}$, define the symbol class $S^k(q^*(T^*X))$ as follows: given a smooth function $\mathscr{A}(x,\xi,z)\in\mathscr{C}^\infty(q^*(T^*X))$ for $\xi\in T^*_xX$ and $z\in q^{-1}(x)\cong N$, it lies in $S^k(q^*(T^*X))$ if and only if for any $j\in\mathbb{N}$, the following norm is finite:
\begin{equation}\label{b11'}
	\lv \mathscr{A}\rv^{(k)}_j=\max_{\begin{subarray}{c}
			\lv\beta\rv,\lv\gamma\rv\leqslant j,\\
			U_\alpha\in\{U_\alpha\}
	\end{subarray}}\sup_{x\in U_\alpha,\xi\in T^*_xX}\langle\xi\rangle^{-k+\lv\gamma\rv}\bv\partial_x^{\beta}\partial_{\xi}^{\gamma}\mathscr{A}(x,\xi,\cdot)\bv_{\mathscr{C}^j(q^{-1}(x))}.
\end{equation}

For $\mathscr{A}\in S^k(q^*(T^*X))$ and $p\in\mathbb{N}$, let $T_{\mathscr{A},p}$ be its \emph{fibrewise Berezin-Toeplitz quantization}, that is, $T_{\mathscr{A},p}(x,\xi)=T_{\mathscr{A}(x,\xi,\cdot),p}$, then $T_{\mathscr{A},p}\in S^k_{F_p}$, indeed, by \eqref{b2}, \eqref{c4} and \eqref{b11'}, we have
\begin{equation}\label{c11}
\lv T_{\mathscr{A},p}\rv^{(k)}_j\leqslant \lv \mathscr{A}\rv^{(k)}_j\ \ \text{for any\ } j\in\mathbb{N}.
\end{equation}

As shown in \eqref{b1'}, the horizontal geodesic flow $(\bm{g}_t)_{t\in\mathbb{R}}$, given in Definition \ref{A4}, acts on functions on $q^*(S^*X)$. Moreover, if we restrict $\mathscr{A}\in S^k(q^*(T^*X))$ to $q^*(S^*X)$, we obtain
\begin{equation}\label{c12}
g_t\cdot T_{\mathscr{A},p}=T_{\bm{g}_t\cdot\mathscr{A},p},
\end{equation}
where the left side is the geodesic flow $(g_t)_{t\in\mathbb{R}}$-action given in \eqref{b1'}.

Now we have the following estimate of the quantum variances.
\begin{prop}\label{Th4.4}
	There is $k\in\mathbb{N}$ such that for any $0<a<b<\infty, 0< T<\infty$, there are $C_{a,b},C_{a,b,T}>0$ such that for $0<h\leqslant1,p\in\mathbb{N}$ and $\mathscr{A}\in S^0(q^*(T^*X))$, we have
\begin{equation}\label{4.12''}
	\begin{split}
		\frac{(2\pi h)^m}{\dim F_p}\sum_{ah^{-2}\leqslant\lambda_{p,j}\leqslant bh^{-2}}\bbv\langle \op_{h_{p,j}}(T_{\mathscr{A},p})u_{p,j},u_{p,j}\rangle_{L^2(X,F_p)}-\int_{S^*X}\overline{\tro}^{\pi^*F_p}[T_{\mathscr{A},p}]d\overline{v}_{S^*X}\bbv^2\\
		\leqslant C_{a,b}\int_{S^*X}\overline{\tro}^{F_p}\Big[T_{\lv\langle\mathscr{A}\rangle_T-\int_{S^*X}\overline{\tro}^{F_p}[T_{\mathscr{A},p}]d\overline{v}_{q^*(S^*X)}\rv_{\mathbb{C}}^2,p}\Big]d\overline{v}_{S^*X}+C_{a,b,T}h\big(\lv\mathscr{A}\rv^{(0)}_k\big)^2.
	\end{split}
\end{equation}
\end{prop}

\begin{pro}
By \eqref{c12}, we get $\langle T_{\mathscr{A},p}\rangle_T=T_{\langle\mathscr{A}\rangle_T,p}$, which implies that
\begin{equation}\label{c15}
	\begin{split}
\langle T_{\mathscr{A},p}\rangle_T-\int_{S^*X}\overline{\tro}^{F_p}[T_{\mathscr{A},p}]d\overline{v}_{q^*(S^*X)}\cdot\mathrm{Id}_{F_p}=T_{\langle\mathscr{A}\rangle_T-\int_{S^*X}\overline{\tro}^{F_p}[T_{\mathscr{A},p}]d\overline{v}_{q^*(S^*X)},p}.
	\end{split}
\end{equation}
According to \eqref{b25}, \eqref{4.7''} and \eqref{c15}, we have
\begin{equation}\label{c16}
	\begin{split}
	\bbV\langle T_{\mathscr{A},p}\rangle_T-&\int_{S^*X}\overline{\tro}^{F_p}[T_{\mathscr{A},p}]d\overline{v}_{q^*(S^*X)}\cdot\mathrm{Id}_{F_p}\bbV_{L^2(S^*X,\mathrm{End}(F_p))}^2\\
		&\leqslant \int_{S^*X}\overline{\tro}^{F_p}\Big[T_{\lv\langle\mathscr{A}\rangle_T-\int_{S^*X}\overline{\tro}^{F_p}[T_{\mathscr{A},p}]d\overline{v}_{q^*(S^*X)}\rv_{\mathbb{C}}^2,p}\Big]d\overline{v}_{S^*X}.
	\end{split}
\end{equation}

Now we simply put $(F,A)=(F_p,T_{\mathscr{A},p})$ in \eqref{b26}, and then apply \eqref{c11}, \eqref{c15} and \eqref{c16} to get \eqref{4.12''}. Note that we are working on an infinite number of bundles $\{F_p\}_{p\in\mathbb{N}}$, so here we should use the \emph{uniformity} of $k,C_{a,b},C_{a,b,T}$ discussed in Remark \ref{B.1}. \qed
\end{pro}

\subsection{Ergodicity of the horizontal geodesic flow}\label{Ce}

We have the following criterion for the ergodicity of the horizontal geodesic flow.
\begin{prop}\label{C6}
	If the geodesic flow $(g_t)_{t\in\mathbb{R}}$ on $S^*X$ is Anosov, and the set of $z\in N$ such that the orbit $\pi_1(X)\cdot z\subset N$ through \eqref{1.8gg} is dense has full measure, then the horizontal geodesic flow $(\bm{g}_t)_{t\in\mathbb{R}}$ on $q^*(S^*X)$ is ergodic with respect to the augmented Liouville measure $dv_{q^*(S^*X)}$. Conversely, if $(\bm{g}_t)_{t\in\mathbb{R}}$ is ergodic with respect to $dv_{q^*(S^*X)}$, the set of $z\in N$ such that the orbit $\pi_1(X)\cdot z\subset N$ is dense has full measure.
\end{prop}

\begin{pro}
	Similar to Proposition \ref{B1'}, it suffices to prove that for any $\mathscr{A}\in\mathscr{C}(q^*(S^*X))$,
	\begin{equation}
		\lim_{T\to\pm\infty}\bbV\langle \mathscr{A}\rangle_T-\int_{q^*(S^*X)}\mathscr{A}d\overline{v}_{q^*(S^*X)}\bbV_{L^2(q^*(S^*X))}^2=0.
	\end{equation}
By the Birkhoff theorem \cite[Theorem 2.3]{MR1377265}, the limit $\mathscr{A}^\pm(x,\xi,z)=\lim_{T\to\pm\infty}\langle \mathscr{A}\rangle_{T}(x,\xi,z)$ exists for almost every $(x,\xi,z)\in q^*(S^*X)$ and we have $\mathscr{A}^+=\mathscr{A}^-$ almost everywhere. We lift all these to $S^*\widetilde{X}\times N$, then for almost every $z\in N$, $\lim_{T\to\pm\infty}\langle \widetilde{\mathscr{A}}\rangle_{T}(\widetilde{x},\widetilde{\xi},z)=\widetilde{\mathscr{A}^\pm}(\widetilde{x},\widetilde{\xi},z)$ for almost every $(\widetilde{x},\widetilde{\xi})\in S^*\widetilde{X}$. Since $\widetilde{\mathscr{A}}$ is the lifting of a \emph{uniformly continuous} function $\mathscr{A}\in\mathscr{C}(q^*(S^*X))$ and $\pi_1(X)$ acts isomorphically on $N$, we see that $\widetilde{\mathscr{A}}$ is a \emph{uniformly continuous} function on $S^*\widetilde{X}\times N$. In particular, for each $z\in N$, the function $\widetilde{\mathscr{A}}(\cdot,z)$ on $S^*\widetilde{X}$ is \emph{uniformly continuous}. 
	
	As in the proof of Proposition \ref{B1'}, for almost every $z\in N$, $\widetilde{\mathscr{A}^\pm}(\cdot,z)$ is a constant for almost every $(\widetilde{x},\widetilde{\xi})\in S^*\widetilde{X}$, and we denote it by $\widetilde{\mathscr{A}^\pm}(z)$. Then $\widetilde{\mathscr{A}^\pm}(\cdot)$ is a  $\pi_1(X)$-invariant measurable function on $N$. Using the Lebesgue density point theorem as in \cite[Proposition 4.2.4]{MR3558990}, the denseness condition in Proposition \ref{C6} implies that the $\pi_1(X)$-action on $N$ is ergodic, so $\widetilde{\mathscr{A}^\pm}(\cdot)$ is a constant almost everywhere as well as $\mathscr{A}^\pm$.
	
	Conversely, if $(\bm{g}_t)_{t\in\mathbb{R}}$ is ergodic, then clearly the $\pi_1(X)$-action on $(N,dv_N)$ is ergodic. We choose a \emph{countable base} of open sets $\{U_i\}_{i\in\mathbb{N}}$ of $N$. If $z\in N$ and $\pi_1(X)\cdot z\subset N$ is not dense, then for some $U_i$, we have $U_i\cap (\pi_1(X)\cdot z)=\emptyset$, equivalently, $z\in N\backslash(\cup_{\gamma\in\pi_1(X)}\gamma^{-1}U_i)$. Since $N\backslash(\cup_{\gamma\in\pi_1(X)}\gamma^{-1}U_i)$ is $\pi_1(X)$-invariant, and $(\cup_{\gamma\in\pi_1(X)}\gamma^{-1}U_i)$ has positive measure, we see that
	$N\backslash(\cup_{\gamma\in\pi_1(X)}\gamma^{-1}U_i)$ is of zero measure.
	\qed
\end{pro}

\begin{remark}
Under the assumption of Proposition \ref{C6}, the horizontal geodesic flow $(\bm{g}_t)_{t\in\mathbb{R}}$ is \emph{partially hyperbolic with center dimension greater than one}. We refer to Pesin \cite{MR2068774} for an introduction to partial hyperbolicity.
\end{remark}

\subsection{Uniform quantum ergodicity}\label{Sdf}

We have the following uniform estimate on the quantum variance.
\begin{theo}\label{D9}
	If the horizontal geodesic flow $(\bm{g}_t)_{t\in\mathbb{R}}$ on $q^*(S^*X)$ is ergodic, then for any $\mathscr{A}\in S^0(q^*(T^*X))$, we have
	\begin{equation}\label{4.23g}
		\begin{split}
			\lim_{h\to0}\sup_{p\in\mathbb{N}}\frac{(2\pi h)^m}{\dim F_{p}}\sum_{ah^{-2}\leqslant\lambda_{p,j}\leqslant bh^{-2}}\bbv\langle \op_{h_{p,j}}(T_{\mathscr{A},p})u_{p,j},&u_{p,j}\rangle_{L^2(X,F_{p})}\\
			&-\int_{q^*(S^*X)}\mathscr{A}d\overline{v}_{q^*(S^*X)}\bbv^2=0.
		\end{split}
	\end{equation}
\end{theo}

\begin{pro}
Recall that the $\pi_1(X)$-action on $(N,L)$, see \eqref{1.8gg}, preserves all the holomorphic and metric structure, then the Bergman kernel $P_p(z,z')$, defined in \eqref{4.1..}, is $\pi_1(X)$-invariant, in other words, for any $\gamma\in \pi_1(X)$ and $(z,z')\in N\times N$, we have
\begin{equation}\label{4.19g}
	\gamma P_p(\gamma^{-1}z,\gamma^{-1}z')\gamma^{-1}=P_p(z,z').
\end{equation}

Since we assume that $(\bm{g}_t)_{t\in\mathbb{R}}$ is ergodic, by Proposition \ref{C6}, we can find $z_0\in N$ such that $\pi_1(X)\cdot z_0\subset N$ is dense, then by \eqref{4.19g}, $P_p(\gamma^{-1}\cdot z_0,\gamma^{-1}\cdot z_0)=P_p(z_0,z_0)$, hence the function $z\in N\mapsto P_p(z,z)$ is \emph{constant}. Since
\begin{equation}
	\begin{split}
		\int_{N}P_p(z,z)dv_{N}=\dim_\mathbb{C} H^{(0,0)}(N,L^p),
	\end{split}
\end{equation}
we get the diagonal Bergman kernel
\begin{equation}
	P_p(z,z)=\frac{\dim_\mathbb{C} H^{(0,0)}(N,L^p)}{\mathrm{Vol}(N)},
\end{equation}
and the following exact version of \eqref{cb5'}:
\begin{equation}\label{4.28}
	\begin{split}
	\overline{\tro}^{H^{(0,0)}(N,L^p)}[T_{\mathcal{H},p}]=\frac{1}{\dim_\mathbb{C} H^{(0,0)}(N,L^p)}\int_{N}\mathcal{H}(z)P_p(z,z)dv_{N}=\int_{N}\mathcal{H}d\overline{v}_{N}.
	\end{split}
\end{equation}

Then from \eqref{4.12''} and \eqref{4.28}, we can get the following estimate \emph{without the $O(p^{-1})$ remainder term}
\begin{equation}\label{5.7''}
	\begin{split}
		\frac{(2\pi h)^m}{\dim F_{p}}\sum_{ah^{-2}\leqslant\lambda_{p,j}\leqslant bh^{-2}}\bbv\langle\op_{h_{p,j}}(T_{\mathscr{A},p})u_{p,j},u_{p,j}\rangle_{L^2(X,F_{p})}-\int_{q_\mu^*(S^*X)}\mathscr{A}d\overline{v}_{q^*(S^*X)}\bbv^2\\
		\leqslant C_{a,b}\bbV\langle \mathscr{A}\rangle_T-\int_{q^*(S^*X)}\mathscr{A}d\overline{v}_{q^*(S^*X)}\bbV^2_{L^2(q^*(S^*X))}+C_{a,b,T}h(\lv \mathscr{A}\rv^{(0)}_k)^2,
	\end{split}
\end{equation}
which is central to our considerations. From \eqref{5.7''}, we easily deduce \eqref{4.23g}.\qed
\end{pro}

Then a uniform version of the diagonal argument leads to the following uniform quantum ergodicity (UQE).
\begin{theo}\label{D10}
	If the horizontal geodesic flow $(\bm{g}_t)_{t\in\mathbb{R}}$ on $q^*(S^*X)$ is ergodic, then there is a subset $\mathbb{B}\subseteq\mathbb{N}^2$ satisfying the uniform density one condition \eqref{c22''} such that, for any $\mathscr{A}\in S^0(q^*(T^*X))$, we have the uniform limit
	\begin{equation}\label{9}
		\lim_{\lambda\to\infty}\sup_{\substack{(p,j)\in\mathbb{B},\\
				\lambda_{p,j}\geqslant \lambda}}\lv\langle \op_{h_{p,j}}(T_{\mathscr{A},p})u_{p,j},u_{p,j}\rangle_{L^2(X,F_{p})}-\int_{q^*(S^*X)}\mathscr{A}d\overline{v}_{q^*(S^*X)}\rv=0.
	\end{equation}
\end{theo}

\begin{pro}
	We take $a=1,b=4$ in \eqref{4.23g}. For $p,r\in\mathbb{N}$, set
	\begin{equation}
		N_{p,r}=\lv\{(p,j)\in\mathbb{N}^2\mid 4^r\leqslant\lambda_{p,j}<4^{r+1}\}\rv.
	\end{equation}
	By Corollary \ref{B5}, there are $r_0\in\mathbb{N}$ and $C>1$ such that
	\begin{equation}\label{c22}
		C^{-1}\leqslant{N_{p,r}}/\big({\dim F_p\cdot 2^{mr}}\big)\leqslant C \ \ \text{for\ } p\in\mathbb{N}, r\geqslant r_0.
	\end{equation}
	This implies that 
	\begin{equation}\label{c23}
		C^{-1}\leqslant\Big(\sum_{j=r_0}^{r}N_{p,j}\Big)/N_{p,r}\leqslant C\ \ \text{for\ }p\in\mathbb{N}, r\geqslant r_0,
	\end{equation}
	
	Now we take a sequence $\{\mathscr{A}_i\}_{i\in\mathbb{N}}$ which is dense in $S^{-\infty}(q^*(T^*X))$ with respect to the uniform norm. Put
	\begin{equation}
		\begin{split}
			\varepsilon_{\ell,r}=\sup_{p\in\mathbb{N}}\frac{1}{N_{p,r}}\sum_{\begin{subarray}{c}0\leqslant i\leqslant\ell,\\
					4^r\leqslant\lambda_{p,j}<4^{r+1}
			\end{subarray}}\bbv\langle \op_{h_{p,j}}(T_{\mathscr{A}_i,p})u_{p,j},u_{p,j}&\rangle_{L^2(X,F_{p})}\\
			&-\int_{q^*(S^*X)}\mathscr{A}d\overline{v}_{q_\mu^*(S^*X)}\bbv^2,
		\end{split}
	\end{equation}
	then for each $\ell\in\mathbb{N}$, we have $\lim_{r\to\infty}\varepsilon_{\ell,r}=0$. We choose a strictly increasing sequence $\{r_\ell\}_{\ell\in\mathbb{N}}$ such that $\varepsilon_{\ell,r}\leqslant 2^{-2\ell}$ for all $r\geqslant r_\ell$.
	
	For $(p,r)\in \mathbb{N}^2$, find the unique $\ell \in\mathbb{N}$ such that $r_\ell\leqslant r<r_{\ell+1}$, then we define a set $J_{p,r}$ as follows: $(p,j)\in\mathbb{N}^2$ is in $J_{p,r}$ if and only if $4^{r}\leqslant\lambda_{p,j}<4^{r+1}$ and 
	\begin{equation}
		\max_{0\leqslant i\leqslant\ell}\bbv\langle \op_{h_{p,j}}(T_{\mathscr{A}_i,p})u_{p,j},u_{p,j}\rangle_{L^2(X,F_{p})}-\int_{q^*(S^*X)}\mathscr{A}_id\overline{v}_{q_\mu^*(S^*X)}\bbv^2\leqslant2^{-\ell}.
	\end{equation}
	Then by the Chebyshev inequality, we have
	\begin{equation}\label{c26}
		\lv\{(p,j)\in\mathbb{N}^2\mid 4^{r}\leqslant\lambda_{p,j}<4^{r+1}\}\backslash J_{p,r}\rv\leqslant 2^\ell\varepsilon_{\ell,r}N_{p,r}\leqslant 2^{-\ell}N_{p,r}.
	\end{equation}
	Now we take
	\begin{equation}
		\mathbb{B}=\bigcup_{p\in\mathbb{N}}\Big(\{(p,j)\mid\lambda_{p,j}<4^{r_1}\}\cup\big(\bigcup_{r\in\mathbb{N}}J_{p,r}\big)\Big).
	\end{equation}
	
	By \eqref{c23} and \eqref{c26}, for any $p\in\mathbb{N}$, when $\lambda\in\mathbb{R}$ satisfies $\log_4\lambda\geqslant r_{2\ell}$, we have
	\begin{equation}\label{4.24}
		\begin{split}
			\bv\{(p,j)\in\mathbb{N}^2\mid\lambda_{p,j}\leqslant\lambda\}\backslash\{(p,j)\in\mathbb{B}\mid\lambda_{p,j}\leqslant\lambda\}\bv&\leqslant\sum_{r=r_0}^{[\log_4\lambda]+1}N_{p,r}2^{-r}\\
			&\leqslant \sum_{r=r_0}^{r_\ell}N_{p,r_\ell}+2^{-\ell}\sum_{r=r_\ell}^{[\log_4\lambda]+1}N_{p,r}\\
			&\leqslant C\cdot2^{-\ell} N_{p,[\log_4\lambda]+1},	
		\end{split}
	\end{equation}
	where the last inequality holds since $[\log_4\lambda]\geqslant\ell+r_\ell$. We have the obvious inequality
	\begin{equation}
		\lv\{(p,j)\in\mathbb{N}^2\mid\lambda_{p,j}\leqslant\lambda\}\rv\geqslant N_{p,[\log_4\lambda]}.
	\end{equation}
	Then from \eqref{c22} and \eqref{4.24} we obtain
	\begin{equation}
		\begin{split}
			\frac{\lv\{(p,j)\in\mathbb{B}\mid\lambda_{p,j}\leqslant\lambda\}\rv}{\lv\{(p,j)\in\mathbb{N}^2\mid\lambda_{p,j}\leqslant\lambda\}\rv}&\geqslant 1-2^{-\ell}C\frac{N_{p,[\log_4\lambda]+1}}{N_{p,[\log_4\lambda]}}\\
			&\geqslant 1-2^{-\ell}C,
		\end{split}
	\end{equation}
	which proves the theorem.\qed
\end{pro}

\subsection{The Borel-Weil-Bott theorem}\label{Cd}

Let $U$ be a compact connected Lie group with Lie algebra $\mathfrak{u}$. For $\mu\in\mathfrak{u}^*$, denote by $\mathcal{O}_\mu=U\cdot\mu$ the orbit of coadjoint action.

We \emph{fix} a maximal torus $T$ of $U$ with Lie algebra $\mathfrak{t}$. Note that if $\mu\in\mathfrak{u}^*$ is regular, we have $\mathcal{O}_\mu\cong U/T$. The integral lattice $\Lambda\subset \mathfrak{t}$ is defined as the kernel of the exponential map $\exp\colon\mathfrak{t}\to T$, and the real weight lattice $\Lambda^*\subset \mathfrak{t}^*$ is defined by $\Lambda^*=\mathrm{Hom}(\Lambda,2\pi\mathbb{Z})$. We choose a set of positive roots $\Phi^+\subset\Lambda^*$, a positive open Weyl Chamber $\mathcal{C}^+$ with closure $\overline{\mathcal{C}^+}$. By the Weyl character formula, finite-dimensional irreducible representations of $U$ are parameterized by $\Lambda^*\cap\overline{\mathcal{C}^+}$.

Set $\mathfrak{r}=[\mathfrak{t},\mathfrak{u}]$, then we have $\mathfrak{u}=\mathfrak{t}\oplus \mathfrak{r}$, and $\mathfrak{u}^*=\mathfrak{t}^*\oplus \mathfrak{r}^*$. Hence we identify $\Lambda^*\cap\overline{\mathcal{C}^+}$ with a subset of $\mathfrak{u}^*$. For $\mu\in \Lambda^*\cap\overline{\mathcal{C}^+}$, let $V_\mu$ be the unique irreducible representation of $U$ with highest weight $\mu$, and let $(L_\mu,h^{L_\mu})$ be the canonical prequantum line bundle on $\mathcal{O}_\mu$. In particular, if $\mu$ is regular, we have $L_\mu=U\times_T\mathbb{C}$, where $T$ acts on $\mathbb{C}$ through $\exp(i\mu)$. We have a natural holomorphic $U$-action on $L_\mu$, which gives a representation of $U$ on $H^{(0,0)}(\mathcal{O}_\mu,L^p_\mu)$ for each $p\in\mathbb{N}$. Then the Borel-Weil-Bott theorem \cite[Theorem \rom{4}]{MR89473} gives an isomorphism
\begin{equation}\label{3.28}
	H^{(0,0)}(\mathcal{O}_\mu,L_\mu^p)\cong V_{p\mu},
\end{equation}
see also \cite[Theorem 8.8]{MR2273508}. Let us give another example similar to \eqref{1.23.}. since $\mathrm{SO}(3)\cong\mathrm{SU}(2)/\{\pm 1\}$, we can take
\begin{equation}\label{4.29}
	(U,\mathcal{O}_\mu,L_\mu,V_{p\mu})=\big(\mathrm{SO}(3),\mathbb{CP}^1,\mathcal{O}(2),\mathrm{Sym}^{2p}(\mathbb{C}^2)\big).
\end{equation}

\begin{remark}
Note that in the proof of Theorem \ref{D9}, the ergodicity of the horizontal geodesic flow $(\bm{g}_t)_{t\in\mathbb{R}}$ imposes a certain symmetry on $(N,L)$, which leads to \eqref{4.28} and \eqref{5.7''}. In the case where $(N,L)=(\mathcal{O}_\mu, L_\mu)$, this symmetry is evident, so \eqref{4.28} and \eqref{5.7''} hold automatically without initially assuming the ergodicity of $(\bm{g}_t)_{t\in\mathbb{R}}$.
\end{remark}

From \eqref{a1.}, \eqref{6} and \eqref{3.28}, we get \eqref{1.16gg}. By Proposition \ref{C6}, if the geodesic flow $(g_t)_{t\in\mathbb{R}}$ on $S^*X$ is Anosov and $\rho(\pi_1(X))\subset U$ is dense, then the horizontal geodesic flow $(\bm{g}_t)_{t\in\mathbb{R}}$ on $q^*(S^*X)$ is ergodic, and we deduce Theorem \ref{C9''}.

\section{Applications and Questions}\label{D}

In this section, we provide further applications of uniform quantum ergodicity (UQE) and discuss questions related to semiclassical measures and potential applications in moduli spaces.

This section is organized as follows. In \cref{Da}, we provide additional examples for main theorems. In \cref{Db}, we present several questions inspired by quantum unique ergodicity.

\subsection{Examples}\label{Da}

First, let us discuss more examples on hyperbolic surfaces similar to \eqref{0.17'}. We use the same notation as in \cref{Cd}.
\subsubsection{Surfaces}
For $X=\Gamma_g\backslash \mathbb{H}^2$, a genus $g$ hyperbolic surface, we have $S^*X\cong\Gamma_g\backslash\mathrm{PSL}(2,\mathbb{R})$ and $\pi_1(X)\cong\Gamma_g\subset \mathrm{PSL}(2,\mathbb{R})$ satisfies
\begin{equation}\label{6.1..}
	\Gamma_g\cong\{a_1,b_1,\cdots,a_g,b_g\mid [a_1,b_1]\cdot[a_2,b_2]\cdots [a_g,b_g]=1\}.
\end{equation}
Recall that $U$ is a compact connected Lie group, if $g\in\mathbb{N}$ is large, then ``almost every" representation $\rho\colon\Gamma_g\to U$ has a dense image: we say that an element $c\in U$ is generic if and only if the closure of the group generated by $c$ is a maximal torus. At the $i$-th step, let $U_{i-1}$ be closure of the group generated by $\{c_j\}_{j=1}^{i-1}$, where $U_0=\emptyset$, and we choose a generic element $c_i\notin U_{i-1}$. Using the Cartan closed subgroup theorem \cite[Theorem 1.3.11]{MR0781344} and counting the dimension of the Lie algebra of $U_i$, it follows that after finite steps, $U_k=U$, in other words, we get a sequence $\{c_j\}_{j=1}^k$ which generates a dense subgroup of $U$. Then any representation $\rho\colon\Gamma_g\to U$ with $\rho(a_j)=c_j$ for $1\leqslant j\leqslant k$ satisfies the denseness condition, for instance, $\rho(b_j)=c_j$ for $1\leqslant j\leqslant k$ and $\rho(a_j)=\rho(b_j)=1$ for $k<j\leqslant g$. By \eqref{3.28}, we can take
\begin{equation}
	(X,N,L,q^*(S^*X),F_p)=\big(\Gamma_g\backslash \mathbb{H}^2,\mathcal{O}_\mu,L_\mu,\Gamma_g\backslash(\mathrm{PSL}(2,\mathbb{R})\times\mathcal{O}_\mu),\Gamma_g\backslash (\mathbb{H}^2\times V_{p\mu})\big).
\end{equation}
For the case of \eqref{4.29} and irrational $\theta\in\mathbb{R}$ in \eqref{0.17'}, put
\begin{equation}\label{d2}
	c_1=\begin{pmatrix}
		\cos\theta\pi       & -\sin\theta\pi &0 \\
		\sin\theta\pi      & \cos\theta\pi   & 0  \\
		0      & 0 &1
	\end{pmatrix},\ \ c_2=\begin{pmatrix}
		1&0&0\\
		0&\cos\theta\pi       & -\sin\theta\pi  \\
		0&\sin\theta\pi      & \cos\theta\pi   \\
	\end{pmatrix},
\end{equation}
by the Euler rotation theorem, $c_1$ and $c_2$ generate a dense subgroup of $\mathrm{SO}(3)$, and the two matrices in \eqref{0.17'} project to the two in \eqref{d2}.

\begin{remark}
		From the Howe-Moore theorem \cite[Theorem 11.2.2]{MR3307755}, we can easily upgrade the ergodicity of the horizontal geodesic flow on $\Gamma_g\backslash(\mathrm{PSL}(2,\mathbb{R})\times N)$ to \emph{mixing}. Moreover, in Mostow rigidity \cite[Theorem 15.1.2]{MR3307755}, the ergodicity of this geodesic flow on $\Gamma_g\backslash\mathrm{PSL}(2,\mathbb{R})$ is translated to the ergodicity of the $\Gamma_g$-action on the product of the circle at infinity $\mathbb{S}^1_\infty\times \mathbb{S}^1_\infty$, from this viewpoint, the ergodicity of the horizontal geodesic flow on $\Gamma_g\backslash(\mathrm{PSL}(2,\mathbb{R})\times N)$ is equivalent to the ergodicity of the $\Gamma_g$-action on $\mathbb{S}^1_\infty\times \mathbb{S}^1_\infty\times N$, then by Proposition \ref{C6}, the $\Gamma_g$-action on $\mathbb{S}^1_\infty\times \mathbb{S}^1_\infty$ is to some extent \emph{weakly mixing}.
		
		Forni-Goldman \cite{MR3931785} proved the ergodicity of the horizontal geodesic flow on $\mathfrak{T}_g\times_{\mathrm{Mod}_g}\mathrm{Rep}(\Gamma_g,\bm{U})$, where $\mathfrak{T}_g$ is the \emph{Teichmüller space}, $\mathrm{Mod}_g$ is the \emph{mapping class group}, and $\mathrm{Rep}(\Gamma_g,\bm{U})$ is the space of representations from $\Gamma_g$ to a compact connected Lie group $\bm{U}$. They use the fact that the $\mathrm{Mod}_g$-action on $\mathrm{Rep}(\Gamma_g,\bm{U})$ is \emph{weakly mixing}, while it is never the case we consider, because $\pi_1(X)$ acts on $N$ isomorphically.

		While any complex structure on $\Gamma_g\backslash \mathbb{H}^2$ induces a complex structure on $\mathrm{Rep}(\Gamma_g,\bm{U})$, the action of $\mathrm{Mod}_g$ does \emph{not} preserve it. Thus, Theorem \ref{D10} cannot be directly applied, but we can still expect associated QE and UQE.
	\end{remark}

\subsubsection{Arithmetic examples}\label{DAB}
Now we consider examples from arithmetic groups. 

Let $K$ be a number field that is totally real with the ring of integers $O_K$. For $k\geqslant 2$, if $\bm{a}_1,\cdots,\bm{a}_k\in O_K$ such that $\bm{a}_i>0,\sigma(\bm{a}_i)<0$ for $1\leqslant i\leqslant k$ and every embedding $\sigma\colon K\to\mathbb{R}$ with $\sigma\neq\mathrm{Id}$ (for instance $K=\mathbb{Q}[\sqrt{2}]$ and all $\bm{a}_i=\sqrt{2}$), we define the associated quadratic forms and special orthogonal groups
\begin{equation}
	\begin{split}
		&a(x)=x_0^2-\bm{a}_1x_1^2-\cdots -\bm{a}_kx_k^2,\ \ \ b(x)=x_0^2-\sigma(\bm{a}_1)x_1^2-\cdots -\sigma(\bm{a}_k)x_k^2,\\
		&G=\mathrm{SO}(a(x),\mathbb{R})\cong\mathrm{SO}(1,k,\mathbb{R}),\ \ \ U=\mathrm{SO}(b(x),\mathbb{R})\cong\mathrm{SO}(1+k,\mathbb{R}).
	\end{split}
\end{equation}
By the Borel density theorem and the Selberg lemma \cite[Corollary 4.5.6, Theorem 4.8.2]{MR3307755}, $G_{O_K}\subset G$ is cocompact, and the image $\sigma(G_{O_K})\subset U$ is dense, and we can pass to a finite index torsion-free subgroup $\Gamma\subset G_{O_K}$ whose image under $\sigma$ is also dense. Then we can take $(X,U,\rho)=(\Gamma\backslash \mathbb{H}^k,\mathrm{SO}(1+k,\mathbb{R}),\sigma)$ for any nonidentity embedding $\sigma$.

\subsection{Semiclassical measures and QUE}\label{Db}

Theorem \ref{B3'} gives equidistribution excluding a \emph{density zero} subset of eigensections. It is natural to consider the question of whether \emph{all} eigensections equidistribute, meaning that we could replace $\mathbb{B}$ with $\mathbb{N}$ in \eqref{3.10}. This is known as \emph{quantum unique ergodicity} (QUE), which was conjectured by Rudnick-Sarnak \cite{MR1266075} for the trivial line bundle over hyperbolic surfaces. On the other hand, Theorem \ref{D10} ensures the \emph{uniform equidistribution} of a \emph{uniformly density one} subset of eigensections. We could also question if we can replace $\mathbb{B}$ with $\mathbb{N}^2$ in \eqref{9}, which we might refer to as \emph{uniform quantum unique ergodicity} (UQUE).

QUE and UQUE also can be stated in terms of the following definition analogous to \cite[Theorem 5.2]{MR2952218}.

In the case of Theorem \ref{B2'}, a \emph{nonnegative functional} $\nu_\mu\colon S^0_{F_\mu}\to\mathbb{C}$ is called a \emph{semiclassical functional} if there are eigensections $\{u_{j_\ell}\}_{\ell\in\mathbb{N}}$ with $\lim_{\ell\to\infty
}\lambda_{j_\ell}=\infty$ and
\begin{equation}\label{5.5}
	\lim_{\ell\to\infty}\langle\op_{h_{j_\ell}}(A)u_{j_\ell},u_{j_\ell}\rangle_{L^2(X,F_{\mu})}=\nu_\mu(A)\ \ \text{for any\ } {A}\in S^0_{F_\mu}.
\end{equation}
In the case of Theorem \ref{D10}, we say that a \emph{nonnegative measure} $\nu_{q^*(T^*X)}$ on $q^*(T^*X)$ is an \emph{augmented semiclassical measure} if there is a series of eigensections $\{u_{p_\ell,j_\ell}\}_{\ell\in\mathbb{N}}$ such that $\lim_{\ell\to\infty
}\lambda_{p_\ell,j_\ell}=\infty$, or equivalently $\lim_{\ell\to\infty}j_\ell/p_{\ell}^{\dim_\mathbb{C}N}=\infty$ by \eqref{b.19.} and \eqref{cb5}, and for any $\mathscr{A}\in S^0(q^*(T^*X))$,
\begin{equation}\label{5.6}
	\lim_{\ell\to\infty}\big\langle\op_{h_{p_\ell,j_\ell}}(T_{\mathscr{A},p_\ell})u_{p_\ell,j_\ell},u_{p_\ell,j_\ell}\big\rangle_{L^2(X,F_{p_\ell})}=\int_{q^*(T^*X)}\mathscr{A}d\nu_{q^*(T^*X)}.
\end{equation}
By nonnegativity we mean that for any $A\in S^0_{F_\mu}$ and $\mathscr{A}\in S^0(q^*(T^*X))$,
\begin{equation}
\nu_\mu\big(A^*A\big)\geqslant0,\ \ \ \nu_{q^*(T^*X)}\big(\lv\mathscr{A}\rv^2\big)\geqslant0.
\end{equation}

Using \eqref{b7}, we have the following weak Gårding inequality \cite[Theorem 4.30]{MR2952218}: for any flat bundle $F$ as in \eqref{2.1} and $\varepsilon>0$, there is $C_\varepsilon>0$ such that for any $A\in S^0_F$ and $0<h\leqslant \min\big\{C_\varepsilon\big(1+\lv A\rv^{(0)}_2\big)^{-2},1\big\}$,
\begin{equation}\label{d7}
\big\langle \op_h(A^*A)u,u\big\rangle_{L^2(X,F)}\geqslant -\varepsilon\lV u\rV^2_{L^2(X,F)},
\end{equation}
and all constants also verify the uniformity discussed in Remark \ref{B.1}.

From \eqref{b.20}, \eqref{b23.}, \eqref{d7} and proceed as in \cite[Theorems 5.2, 5.3, 5.4]{MR2952218}, we get the following reult.
\begin{prop}\label{E.2}
For any sequence $\{u_{j_\ell}\}_{\ell\in\mathbb{N}}$ of eigensections of $\Delta^{F_\mu}$ with $\lim_{\ell\to\infty}\lambda_{j_\ell}=\infty$, there exists a subsequence which is weak-star convergent to a nonnegative functional $\nu_{\mu}$. Furthermore, every semiclassical functional $\nu_\mu$ satisfies $\mathrm{supp}(\nu_\mu)\subseteq S^*X$ and it is invariant with respect to the geodesic flow, that is, for any $A\in\mathscr{C}^\infty(S^*X,\pi^*\mathrm{End}(F_\mu))$,  we have $\nu_\mu\big(g_t\cdot A\big)=\nu_\mu\big(A\big)$.

Similarly, for any sequence $\{u_{p_\ell,j_\ell}\}_{\ell\in\mathbb{N}}$ of eigensections of $\{\Delta^{F_{p_\ell}}\}$ with $\lim_{\ell\to\infty
}\lambda_{p_\ell,j_\ell}=\infty$, there is a subsequence that is weak-star convergent to a nonnegative probability measure. Moreover, every augmented semiclassical measure is supported on $q^*(S^*X)$ and is invariant under the horizontal geodesic flow. Therefore, we denote it by $\nu_{q^*(S^*X)}$, and for any $\mathscr{A}\in\mathscr{C}^\infty(q^*(S^*X))$, we have $\nu_{q^*(S^*X)}\big(\bm{g}_t\cdot \mathscr{A}\big)=\nu_{q^*(S^*X)}\big(\mathscr{A}\big)$.
\end{prop}
Note that in the proof of the second part of Proposition \ref{E.2}, we should use \eqref{c11} and \eqref{d7} to set an upper bound for $h$ uniformly for $p\in\mathbb{N}$.

Recall the \emph{Liouville functional} $A\in S_{F_\mu}^0\mapsto \int_{S^*X}\overline{\tro}^{\pi^*F_\mu}[A]d\overline{v}_{S^*X}$ and the \emph{augmented Liouville measure} $d\overline{v}_{q^*(S^*X)}$. Then the QUE conjecture states that the only semiclassical functional is the Liouville functional, while the UQUE conjecture states that the only augmented semiclassical measure is the augmented Liouville measure.

\subsubsection{Arithmetic UQUE}

If $X$ is an arithmetic surface and $F=\mathbb{C}$, there are extra symmetries called \emph{Hecke operators}, which commute with each other and the Laplacian. The QUE has been proved by Lindenstrauss \cite{MR2195133} for eigenfunctions $\{u_j\}_{j\in\mathbb{N}}$ of the Laplacian which are also eigenfunctions of all Hecke operators when $X$ is compact.

When $(X,F_\mu\ (\text{or\ }F_{p}))=\big(G_{O_K}\backslash\mathbb{H}^2,G_{O_K}\backslash(\mathbb{H}^2\times\mathrm{Sym}^{2p}(\mathbb{C}^2))\big)$ as in \cref{DAB}, there are also Hecke symmetries on $L^2\big(G_{O_K}\backslash\mathrm{PSL}(2,\mathbb{R}),G_{O_K}\backslash(\mathrm{PSL}(2,\mathbb{R})\times\mathrm{Sym}^{2p}(\mathbb{C}^2))\big)$. Therefore, we may ask if the arithmetic UQUE hold for the cases we considered.
\subsubsection{The entropy of semiclassical measures}

Anantharaman \cite{MR2434883} gave a restriction on the set of semiclassical measures: if $X$ has Anosov geodesic flow and $F=\mathbb{C}$, then the Kolmogorov-Sinai entropy $h_{\mathrm{KS}}(\nu)$ of any semiclassical measure $\nu$, see \cite[(9.1.17)]{MR3558990} for a definition, satisfies $h_{\mathrm{KS}}(\nu)>0$.

Restricting the symbol classes $S^0_{F_\mu}, S^0(q^*(T^*X))$ to scalar symbols $S^0_{\mathbb{C}}$ in \eqref{5.5} and \eqref{5.6}, we get the property of pushforward measures $q_*(\nu_\mu),q_*(\nu_{q^*(S^*X)})$ on $S^*X$. 

Therefore, we hope that after a minor modification in \cite{MR2434883}, we can prove that, in the case of Theorem \ref{B2'}, or the special case of Theorem \ref{D10} satisfying Proposition \ref{C6}, the entropy of $q_*(\nu_\mu)$ and $q_*(\nu_{q^*(S^*X)})$ for any semiclassical functional $\nu_\mu$ and measure $\nu_{q^*(S^*X)}$ is still positive. Consequently, we shall have $h_{\mathrm{KS}}(\nu_{q^*(S^*X)})\geqslant h_{\mathrm{KS}}(q_*(\nu_{q^*(S^*X)}))>0$.

\subsubsection{The support of semiclassical measures}

Dyatlov-Jin-Nonnenmacher \cite{MR4374954} showed a different kind of restriction when $X$ is a surface with Anosov geodesic flow and $F=\mathbb{C}$. For each semiclassical measure $\nu$ on $S^*X$, we have $\mathrm{supp}(\nu)=S^*X$, namely, $\nu_{}(U)>0$ for any nonempty open set $U\subseteq S^*X$, see also Dyatlov-Jin \cite{MR3849286} for the constant sectional curvature case. By extending their main estimate \cite[Theorem 2]{MR4374954} to scalar symbols $S^0_\mathbb{C}\subset S^0_F$ for any flat bundle $F$ and ensuring uniformity of constants as in Remark \ref{B.1}, we can prove analogous results for semiclassical functionals and augmented measures, using the invariance with respect to the geodesic flow or the horizontal geodesic flow in Proposition \ref{E.2}.

\addcontentsline{toc}{section}{References}
\bibliographystyle{abbrv}

\end{document}